\documentclass[smallextended, useAMS, usenatbib]{svjour3}
\usepackage[breaklinks,colorlinks,urlcolor=blue,citecolor=blue,linkcolor=blue]{hyperref}
\usepackage{mathrsfs}
\usepackage{graphicx}
\usepackage{amsmath, amssymb}
\usepackage{subfigure}
\usepackage{epsfig}
\usepackage{enumerate}
\usepackage[numbers]{natbib}

\usepackage{color}

\newcommand       \be          {\begin{eqnarray}}
\newcommand       \ee          {\end{eqnarray}}

\begin{document}

\title{Stellar Tidal Disruption Events in General Relativity}

\author{Nicholas C.~Stone$^1$
\and Michael Kesden$^2$
\and Roseanne M.~Cheng$^3$
\and Sjoert van Velzen$^4$}

\institute{
$^1$Columbia Astrophysics Laboratory, Columbia University, New York, New York, 10027, USA
\and
$^2$Department of Physics, The University of Texas at Dallas, 800 W. Campbell Rd., Richardson, Texas 75080, USA
\and
$^3$Computational Physics and Methods Group, Los Alamos National Laboratory, P.O. Box 1663, Los Alamos, New Mexico, 87545, USA
\and
$^4$Center for Cosmology and Particle Physics, New York University, New York, New York, 10003, USA
}

\date{Received: date / Accepted: date}

\maketitle

\begin{abstract}
A tidal disruption event (TDE) ensues when a star passes too close to the supermassive black hole (SMBH) in a galactic center and is ripped apart by the tidal field of the SMBH.  The gaseous debris produced in a TDE can power a bright electromagnetic flare as it is accreted by the SMBH; so far, several dozen TDE candidates have been observed.  For SMBHs with masses above $\sim 10^7 M_\odot$, the tidal disruption of solar-type stars occurs within ten gravitational radii of the SMBH, implying that general relativity (GR) is needed to describe gravity.  Three promising signatures of GR in TDEs are: (1) a super-exponential cutoff in the volumetric TDE rate for SMBH masses above $\sim 10^8 M_\odot$ due to direct capture of tidal debris by the event horizon, (2) delays in accretion disk formation (and a consequent alteration of the early-time light curve) caused by the effects of relativistic precession on stream circularization, and (3) quasi-periodic modulation of X-ray emission due to global precession of misaligned accretion disks and the jets they launch.  We review theoretical models and simulations of TDEs in Newtonian gravity, then describe how relativistic modifications give rise to these proposed observational signatures, as well as more speculative effects of GR.  We conclude with a brief summary of TDE observations and the extent to which they show indications of these predicted relativistic signatures.
\end{abstract}

\section{Introduction}
\label{sec:intro}

In classical general relativity (GR), black holes are simple objects fully characterized\footnote{The third parameter of the general Kerr-Newman metric, electric charge, is expected to be effectively zero for any astrophysical black hole.} by their mass $M_\bullet$ and dimensionless spin magnitude $\chi_\bullet = cS/GM_\bullet^2 < 1$, where $c$ is the speed of light, $S$ is the rotational angular momentum of the black hole, and $G$ is Newton's gravitational constant \cite{1963PhRvL..11..237K, 1971PhRvL..26..331C}.  If the masses and spins of astrophysical black holes can be measured, one can calculate their properties and search for deviations from the predictions of GR.  Because GR is a scale-free theory, black holes may exist with arbitrary masses, but astrophysical black holes have been discovered in two distinct mass regimes.  Stellar-mass black holes ($5 M_\odot \lesssim M_\bullet \lesssim 50 M_\odot$ \cite{Belczynski+16}) are believed to form in the collapse of massive stars \cite{2003ApJ...591..288H} and were first observed from the X-rays they emit during accretion from binary stellar companions \cite{1972Natur.235...37W, 1972Natur.235..271B}.  The merger of two stellar-mass black holes was the source of the first gravitational waves (GWs) observed by the Laser Interferometer Gravitational-wave Observatory (LIGO) \cite{2016PhRvL.116f1102A}.  The masses and spins of both the merging black holes and final black hole produced were measured in this event \cite{2016PhRvL.116x1102A}, and in subsequently discovered black-hole mergers.  The measured mass and spin values, as well as other parameters of the gravitational waveforms, were found to be consistent with the predictions of GR \cite{2016PhRvL.116v1101A}, constraining some but by no means all modified theories of gravity.  No GWs have yet been observed from the second known class of astrophysical black holes, the supermassive black holes (SMBHs) with masses $10^6 M_\odot \lesssim M_\bullet \lesssim 10^{10} M_\odot$.  However, the masses and spin of these SMBHs may still be measured through traditional electromagnetic observations.

SMBHs are known to lurk in the nuclei of most galaxies comparable to or greater than\footnote{The presence of massive black holes in the nuclei of smaller galaxies is poorly constrained due to instrumental limitations.} the Milky Way in size \cite{1995ARA&A..33..581K, 1998AJ....115.2285M, Ferrarese&Ford05, Graham&Spitler09, Kormendy&Ho13}.  A small fraction \cite{2003MNRAS.346.1055K} of these SMBHs, with mass accretion rates $\dot{M} \gtrsim 10^{-2} \dot{M}_{\rm Edd}$, are considered to be ``active.''  The Eddington accretion rate
\begin{equation}\label{eq:edd}
\dot{M}_{\rm Edd} = \frac{4\pi GM_\bullet m_p}{\eta c \sigma_T} \simeq 2.2\times10^{-2}M_\odot~{\rm yr}^{-1} \left( \frac{\eta}{0.1} \right)^{-1} \left( \frac{M_\bullet}{10^6M_\odot} \right)
\end{equation}
is an estimate for the maximum rate at which a SMBH can accrete gas, where $m_p$ is the proton mass, $\sigma_T$ is the Thomson scattering cross section, $\eta<1$ is a dimensionless radiative efficiency factor, and we have assumed fully ionized hydrogen.  Although this is an idealized limit, and a number of theoretical mechanisms have been proposed to circumvent it \cite{Abramowicz+88, Watarai+01, Jiang+14}, observational evidence suggests it is obeyed \cite{Heckman+04} by the SMBHs powering most active galactic nuclei (AGN). 

AGN are among the brightest astrophysical sources of thermal and nonthermal radiation, and their extravagant emission can easily be observed across the electromagnetic spectrum, sometimes to cosmological distances \cite{Fan06}.  The masses \cite{Peterson93} and spins \cite{Reynolds13} of SMBHs in AGN can often be inferred, albeit with significant uncertainty, by modeling the electromagnetic spectra of the presumably steady-state accretion disks and outflows that surround them.  While SMBH masses can be measured with some precision \cite{Greenhill+95, Humphreys+13}, existing spin measurement techniques are more inexact and suffer from notable systematic uncertainties \cite{Reynolds&Fabian08}.  The majority of SMBHs, however, are ``quiescent'' and accrete at a more modest rate, making them less luminous and quite challenging to observe directly.  While it may be possible to directly measure the mass and spin of the two SMBHs which subtend the largest solid angle on the sky (Sgr A$^\star$, the SMBH in the center of the Milky Way, is one of these, and the SMBH in the giant elliptical galaxy M87 is the other) through very long baseline interferometry \cite{Goddi17}, the prospects for extending this technique to other quiescent SMBHs appear dim.  The masses of a few dozen of the closest quiescent SMBHs can be inferred dynamically, through their imprint on the orbits of stars or cold circumnuclear gas \cite{Kormendy&Ho13}, but their spins leave no such dynamical imprint on scales realistically resolvable with electromagnetic instruments.  

Aspiring SMBH demographers are thus left with a painful dilemma: the majority of astrophysical SMBHs exist at any given time in a quiescent state, but with a handful of exceptions, there is no obvious way to measure their masses, let alone spins.  One possible tool to resolve this dilemma, which is the subject of this review, is observation of tidal disruption events (TDEs). These events mark the destruction of a star, of mass $M_\star$ and radius $R_\star$, that passes within the tidal radius of an SMBH:
\begin{equation} \label{eq:rt}
r_t \equiv R_\star \left( \frac{M_\bullet}{M_\star} \right)^{1/3} \simeq 2.3 \times 10^{-6}\,{\rm pc} \left( \frac{R_\star}{R_\odot} \right) \left( \frac{M_\star}{M_\odot} \right)^{-1/3} \left( \frac{M_\bullet}{10^6\,M_\odot} \right)^{1/3}.
\end{equation}
This is the approximate radius at which the star's self gravity is overwhelmed by the tidal field of the SMBH \cite{Wheeler71, Hills75}; to derive it, we have equated the Newtonian surface tidal acceleration $GM_\bullet R_\star/r_t^3$ with the star's self-gravitational acceleration $GM_\star / R_\star^2$.  About half of the stellar debris becomes gravitationally bound to the SMBH and can power a luminous, multi-wavelength flare \cite{Lidskii&Ozernoi79, Rees88}.  Although the tidal disruption process is reasonably well understood from a theoretical perspective, the hydrodynamics of the tidal debris, its subsequent accretion by the SMBH, and the radiation emitted during the various stages of this process all remain fairly uncertain.  However, by comparing $r_t$ to the gravitational radius $r_g = GM_\bullet/c^2$, it is easy to see that many TDEs are fundamentally {\it general relativistic} events.  While the tidal radius is modestly non-relativistic for the smallest SMBHs, $r_t \sim r_g$ for a wide range of parameter space.  

If we define, somewhat arbitrarily, $r_t \le 10 r_g$ as the threshold separating ``relativistic'' from ``non-relativistic'' TDEs, we find a SMBH mass 
\begin{equation} \label{E:MGR}
M_{GR} = M_\star^{-1/2} \left( \frac{c^2 R_\star}{10G} \right)^{3/2} \simeq 1\times 10^7 M_\odot \left( \frac{R_\star}{R_\odot} \right)^{3/2} \left( \frac{M_\star}{M_\odot} \right)^{-1/2},
\end{equation}
above which we may infer that GR is fundamental to understanding TDE physics.  The above $M_{GR}$ is heuristic, but points towards a better-known calculation with much clearer physical significance.  Equating $r_t$ with the Schwarzschild radius, $2r_g$ (the approximate size of a black-hole event horizon), it can be seen that TDEs are hidden from the outside universe for SMBHs larger than a ``Hills mass'' \cite{Hills75}
\begin{equation}
M_{H} = M_\star^{-1/2} \left( \frac{c^2 R_\star}{2G} \right)^{3/2} \simeq 1\times 10^8 M_\odot \left( \frac{R_\star}{R_\odot} \right)^{3/2} \left( \frac{M_\star}{M_\odot} \right)^{-1/2}. \label{eq:MHills}
\end{equation}
SMBHs with mass $M_\bullet \gtrsim M_{H}$ swallow stars whole, leaving no gas outside the event horizon to produce an electromagnetically luminous flare.  Note that this Hills mass implies that only an intermediate-mass black hole (IMBH) with $M_\bullet < 10^5 M_\odot$ can disrupt a white dwarf ($M_\star \simeq 0.6 M_\odot, R_\star \simeq 10^{-2} R_\odot$), while even the most massive SMBH ($M_\bullet \simeq 10^{10} M_\odot$) can easily disrupt a star on the tip of the red-giant branch ($M_\star \simeq M_\odot, R_\star \simeq 10^2 R_\odot$) \cite{MacLeod+12}.  This estimate of the Hills mass is gauge-dependent and quite approximate; later in this review we will define the threshold for direct capture more precisely.

In the future, TDEs may become a powerful tool for measuring SMBH demography (masses and spins).  More speculatively, the light curves and spectral-energy distributions of individual flares could perhaps be used to test alternatives to standard black-hole solutions \cite{Lu+17} or to perform consistency checks on aspects of classical GR such as the no-hair theorem (\S \ref{sec:circularization}).  It is important to note, however, that there are many open questions concerning TDE hydrodynamics and radiative processes, and at least some of these questions must be resolved before tidal disruption flares can reliably probe SMBH demography, let alone GR.  In this review, we will emphasize, wherever possible, which aspects of TDE physics are well understood and which remain elusive.  

The general outline of this review is as follows.  In \S \ref{sec:basic}, we give a brief overview of tidal disruption in Newtonian gravity, treating the SMBH as a Keplerian point particle of mass $M_\bullet$.  In \S \ref{sec:NewtonianAccretion}, we will discuss how the gaseous debris produced following tidal disruption is accreted by the SMBH in Newtonian gravity.  In \S \ref{sec:rates}, we describe how stellar diffusion into the phase space ``loss cone'' determines the TDE rate as a function of SMBH mass and properties of the host galaxy.  In \S \ref{sec:GR1}, we present relatively well understood ways in which GR may alter the process of tidal disruption and discuss how these may leave observable imprints in current and future TDE samples.  In \S \ref{sec:GR2}, we discuss the more uncertain ways in which GR can mediate the assembly and behavior of TDE accretion disks.  In \S \ref{sec:obs}, we review existing astronomical observations of TDEs, connecting these observations, when possible, back to the relativistic physics of earlier sections.  We summarize in \S \ref{sec:disc}.

\section{Tidal Disruption Events in Newtonian Gravity}\label{sec:basic}

The disruption of a star involves many types of physics, but the dominant effects are due to gravity and hydrodynamics.  A variety of techniques may be employed to study the disruption process by which the initially self-bound star becomes tidal debris.  Analytic or semi-analytic models either tackle the gravity alone, or combine gravity with hydrodynamics in a highly simplified way.  Numerical hydrodynamic simulations are usually needed for precision calculations, but as we shall see in this section, these simulations agree to leading order with simpler models.  General-relativistic gravity is essential for understanding TDEs around SMBHs with $M_\bullet \gtrsim M_{GR}$, and as we shall see, it may even be important for TDEs around smaller SMBHs. However, we will gain some preliminary intuition by considering tidal disruption in Newtonian gravity.  In this section, we shall consider gravity in a purely Newtonian fashion, and treat the SMBH as a point mass with a perfectly absorbing boundary layer at radius $r=2r_g$.

\subsection{Stellar Tidal Disruption}
\label{SS:NewtTDE}

\begin{figure}
{ \begin{center}
\includegraphics[scale=0.45]{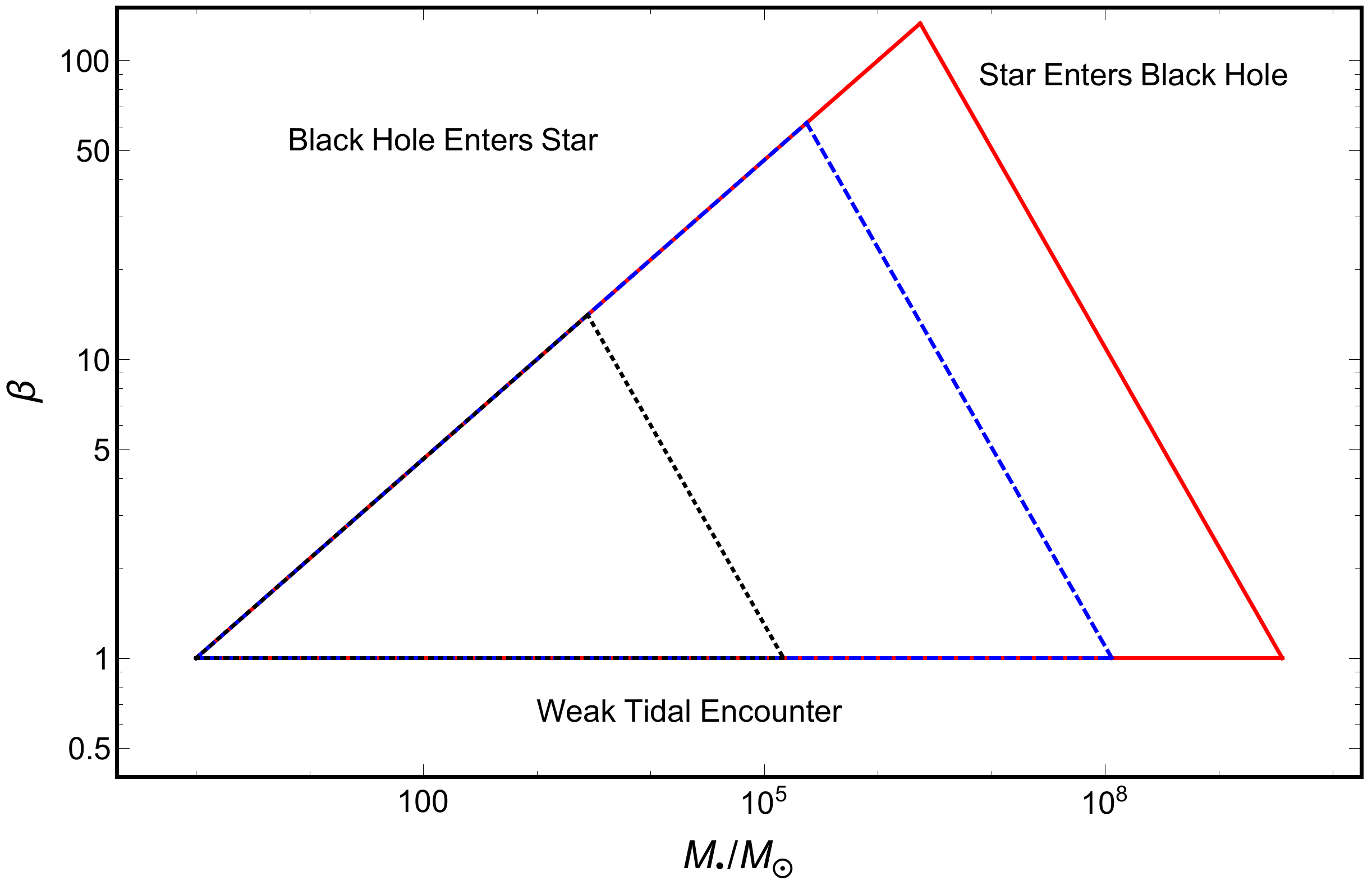}
\end{center}
\caption{\label{figure:TDE_parameters}
The ``TDE Triangle'' identifying the region of the $\beta-M_\bullet$ plane in which Newtonian tidal disruptions are permitted.  In the upper left of diagram, tidal disruption is prevented by physical collisions (the black hole penetrates inside the star, $r_p < R_\star, 2r_g < R_\star$).  In the upper right, stars plunge into the event horizon ($r_p < 2r_g, R_\star < 2r_g$).  The blue dashed lines bound the permitted parameter space for TDEs of Sun-like stars ($M_\star = M_\odot$, $R_\star = R_\odot$), which cannot achieve penetration factors above $\beta \simeq 50$.  The black dotted triangle corresponds to a white dwarf ($M_\star = 0.6 M_\odot$, $R_\star=0.01R_\odot$); typical white dwarfs can only be disrupted by intermediate-mass black holes, not SMBHs.  The red solid triangle is for a red giant star ($M_\star = M_\odot$, $R_\star = 10R_\odot$).  The large Hills masses $M_{\rm H}$ of these stars permit disruptions by almost any known SMBH.  Adapted from Ref. \cite{StoneThesis}.
}  }
\end{figure}

Let us begin by considering a victim star, of mass $M_\star$ and radius $R_\star$, approaching a SMBH with mass $M_\bullet \gg M_\star$.  Tidally disrupted stars are typically on orbits with apocenter distances $r_a$ comparable to the SMBH influence radius
\begin{equation}
r_h \simeq \frac{GM_\bullet}{\sigma^2} = 0.43\,{\rm pc} \left( \frac{M_\bullet}{10^6\,M_\odot} \right) \left( \frac{\sigma}{100\,{\rm km/s}} \right)^{-2},
\end{equation}
where $\sigma$ is the velocity dispersion of the host galaxy of the SMBH.  As $r_a \sim r_h \gg r_t$, the star's orbit is very nearly parabolic, with typical eccentricity $e_o$ given by
\begin{equation} \label{E:eccstar}
1 - e_o \lesssim \frac{r_t}{r_h} \simeq 5.3 \times 10^{-6} \left( \frac{R_\star}{R_\odot} \right) \left( \frac{M_\star}{M_\odot} \right)^{-1/3} \left( \frac{M_\bullet}{10^6\,M_\odot} \right)^{-2/3} \left( \frac{\sigma}{100\,{\rm km/s}} \right)^2,
\end{equation}
though the eccentricity may be even higher if the orbital pericenter $r_p < r_t$.  The specific orbital energy $\epsilon_o \sim GM_\bullet/r_h \simeq \sigma^2,$ and the specific angular momentum $L_o = (2GM_\bullet r_p)^{1/2}$.  The strength of the disruption can be quantified by the penetration factor\footnote{The strength of the tidal encounter is sometimes quantified with the alternative dimensionless impact parameter $\eta_t \equiv (M_\star/M_\bullet)^{1/2}(r_p/R_\star)^{3/2}= \beta^{-3/2}$.}
\begin{equation} \label{E:beta}
\beta \equiv \left( \frac{L_t}{L_o} \right)^2 = \frac{r_t}{r_p}\,,
\end{equation}
where $L_t = (2GM_\bullet r_t)^{1/2}$.  Not all values of $\beta$ are accessible; very high-$\beta$ encounters will be prevented by physical collisions, where either the SMBH enters the star ($2r_g < R_\star$) or the star plunges directly through the event horizon ($R_\star < 2r_g)$.  We illustrate this in Fig.~\ref{figure:TDE_parameters}, which shows the ``TDE Triangle'' outlining geometrically accessible $\beta$ values \cite{Luminet&Pichon89}.

We earlier defined $r_t$ in Eq.~(\ref{eq:rt}) by equating the Newtonian tidal acceleration to that of self-gravity on the surface of our victim star.  This definition ignores the hydrodynamic details of the disruption process; in reality, full disruption only occurs for penetration factors $\beta>\beta_d$, a threshold value of order unity that demarcates the boundary between full and partial disruption \cite{Diener+97}.  There will be a second threshold, $\beta_p$, for partial disruption; stars on orbits with $\beta_p < \beta < \beta_d$ will only have their outer layers tidally stripped, while a high-density stellar core remains intact \cite{Ivanov&Novikov01,Bogdanovic&Cheng+14}.  Stars on orbits with $\beta < \beta_p$ survive their encounter with the SMBH fully intact, though tides may transfer orbital energy into the star by exciting normal modes \cite{Alexander&Morris03}.  
In Newtonian gravity, the values of $\beta_d$ and $\beta_p$ depend only on the internal structure (i.e. density profile) of the victim star.  Their precise values have been quantified in Newtonian hydrodynamical simulations, at least for the limiting cases of polytropic stars with density profiles determined by the Lane-Emden equation.  Simulations performed using a range of modern Lagrangian techniques agree that for $\gamma=5/3$ polytropic models (appropriate for low-mass main-sequence stars and white dwarfs), $\beta_d \approx 0.92$, while for more centrally concentrated $\gamma=4/3$ models (appropriate for intermediate-mass main-sequence stars and very high-mass white dwarfs), $\beta_d \approx 2.01$ \cite{Mainetti+17}.  These results are in good agreement with earlier hydrodynamical simulations using an Eulerian code \cite{Guillochon&RamirezRuiz13}.  The threshold for full disruption is higher in giant-branch stars, which have a much greater density contrast between their core and envelope \cite{MacLeod+12}.  While the Newtonian disruption thresholds for polytropic stellar models are well understood, it is worth noting that $\beta_d$ is not known in Newtonian gravity for fully realistic stellar-structure models, nor is the polytropic $\beta_d$ as precisely measured in fully relativistic gravity\footnote{Although see Refs. \cite{Diener+97, Cheng&Evans13}.}.

Once the star has been disrupted, elements of the tidal debris move on nearly geodesic orbits about the SMBH.  This state of tidal free fall is often modelled analytically using the ``frozen-in'' approximation, in which one assumes that the star is a static sphere which shatters into pieces (i.e. instantaneous loss of self-gravity) when it reaches the tidal radius.  In this approximation, one can estimate the spread in specific energy of the tidal debris by assuming that each fluid element retains the velocity of the star prior to disruption, but acquires a different gravitational potential energy corresponding to its distance from the SMBH.  Taylor expanding the Newtonian potential of the SMBH about the tidal radius $r_t$ and assuming the victim star maintains its unperturbed radius $R_\star$ at disruption, we find that the most bound element of the tidal debris has a specific binding energy \citep{Rees88}
\begin{equation}
\Delta\epsilon \simeq \frac{GM_\bullet R_\star}{r_t^2} = \epsilon_\star \left( \frac{M_\bullet}{M_\star} \right)^{1/3}, \label{eq:energy}
\end{equation}
where $\epsilon_\star \equiv GM_\star/R_\star$ is an estimate of the star's self-binding energy.  The portion of the tidal debris bound to the SMBH has a range of specific binding energies $-\Delta\epsilon \lesssim \epsilon \leq 0$; the inequality $\Delta\epsilon \gg \epsilon_\star$ supports the neglect of the star's internal forces in the ``frozen-in'' approximation.  While this approximation is crude, it is in reasonable agreement with detailed hydrodynamical simulations of the disruption process \citep{Guillochon&RamirezRuiz13, Lodato+09}.  It is worth noting that many past works overestimated the frozen-in energy spread as $GM_\bullet R_\star/r_p^2$.  While this estimate would be appropriate if the tidal ``freezing'' occurred at pericenter $r_p$ rather than the tidal radius $r_t$, it is incorrect because the star's self gravity cannot hold the star together against the SMBH tidal force for distances $r < r_t$ \cite{Stone+13a}.

Working within the frozen-in approximation, we can use the TDE specific energy hierarchy to determine the fate of the tidal debris.  First, because $\Delta \epsilon \gg \epsilon_o$, almost exactly $50\%$ of the tidal debris is unbound from the SMBH during disruption and flies off to infinity on hyperbolic orbits.  The other $\approx 50\%$ becomes more tightly bound to the SMBH but remains on highly elliptical orbits with a minimum eccentricity
\begin{equation} \label{E:emin}
1 - e_{\rm min} \simeq \frac{\Delta \epsilon L_t^2}{\beta(GM_\bullet)^2} = \frac{2}{\beta} \left( \frac{M_\star}{M_\bullet} \right)^{1/3} \ll 1~.
\end{equation}
For the bound tidal debris to evolve into a traditional accretion disk (i.e. one whose gas flows on quasi-circular orbits), it must circularize by increasing the magnitude of its specific binding energy to
\begin{equation} \label{E:Ec}
\epsilon_c = \frac{GM_\bullet}{2r_c} =  \frac{\beta}{4} \left( \frac{M_\bullet}{M_\star} \right)^{1/3}\Delta\epsilon \gg \Delta\epsilon,
\end{equation}
where $r_c = 2r_t/\beta$ is the circularization radius for an orbit with penetration factor $\beta$.  The manner in which tidal debris circularizes (and whether it manages to do so at all) is a major open question.  We shall return to the problem of debris circularization in Newtonian gravity in \S \ref{sec:NewtonianCircularization} and in GR in \S \ref{sec:circularization}.

If the disruption process yields tidal debris with a mass per unit specific binding energy distribution $dM/d\epsilon$, the rate at which mass falls back to pericenter for the first time will be
\begin{equation} \label{E:fallback}
\dot{M}_{\rm fall} = \frac{dM}{d\epsilon} \left| \frac{d\epsilon}{dt} \right| = \frac{dM}{d\epsilon} \frac{(2\pi GM_\bullet)^{2/3}}{3} t^{-5/3}
\end{equation}
for $t > t_{\rm min}$, where
\begin{equation}
t_{\rm min} = 2\pi GM_\bullet (2\Delta \epsilon)^{-3/2} \simeq 41\,{\rm d} \left( \frac{R_\star}{R_\odot} \right)^{3/2} \left( \frac{M_\star}{M_\odot} \right)^{-1} \left( \frac{M_\bullet}{10^6\,M_\odot} \right)^{1/2}
\end{equation}
is an estimate of the orbital period of the most tightly bound tidal debris.  Approximating the differential mass distribution as a top-hat $dM/d\epsilon = M_\star / (2\Delta \epsilon)$ between $-\Delta \epsilon$ and $+\Delta \epsilon$, we find that the Eddington ratio at $t=t_{\rm min}$ is
\begin{equation}
    \frac{\dot{M}_{\rm peak}}{\dot{M}_{\rm Edd}}\simeq 130 \left(\frac{\eta}{0.1} \right) \left( \frac{R_\star}{R_\odot} \right)^{-3/2} \left( \frac{M_\star}{M_\odot} \right)^{2} \left( \frac{M_\bullet}{10^6\,M_\odot} \right)^{-3/2},    \label{eq:MDotNormed}
\end{equation}
and the time it takes for accretion to become sub-Eddington (for cases where $\dot{M}_{\rm peak} > \dot{M}_{\rm Edd}$) is
\begin{equation}
    t_{\rm Edd}\simeq 760~{\rm d}~ \left(\frac{\eta}{0.1} \right)^{3/5} \left( \frac{R_\star}{R_\odot} \right)^{3/5} \left( \frac{M_\star}{M_\odot} \right)^{1/5} \left( \frac{M_\bullet}{10^6\,M_\odot} \right)^{-2/5}.   \label{eq:tEdd}
\end{equation}
The super-Eddington accretion rates that characterize the early periods of mass fallback in TDEs are likely conducive to the launching of relativistic jets \cite{Giannios&Metzger11}, although this is probably a necessary but not sufficient condition for the launching of powerful jets \cite{Tchekhovskoy+14}.

The assumption of a top-hat debris-mass distribution is simplistic, but the scaling relations in Eqs.~(\ref{eq:MDotNormed}) and (\ref{eq:tEdd}) are in good agreement with detailed hydrodynamical simulations, which provide more precise calibrations of the numerical prefactors on $\dot{M}_{\rm peak}$ and $t_{\rm Edd}$ \cite{Guillochon&RamirezRuiz13}.  Physical $dM/d\epsilon$ distributions are shaped by both the internal density profile of the victim star \cite{Lodato+09} and, more weakly, by the $\beta$ parameter \cite{Guillochon&RamirezRuiz13}.  However, $dM/d\epsilon$ does achieve a constant value for $|\epsilon| \ll \Delta \epsilon$, implying that $\dot{M} \propto t^{-5/3}$ at late times\footnote{Although the late-time power law index can be somewhat shallower or steeper for partial disruptions \cite{Guillochon&RamirezRuiz13}.}.  The general accuracy of the analytic estimates in this section arises, in part, from the hierarchy of specific energies in the TDE problem: $\epsilon_c \gg \Delta \epsilon \gg \epsilon_\star \gtrsim \epsilon_o$.  While there are special cases in which this hierarchy breaks down, it is usually robust for astrophysical TDEs.

Early work on TDEs generally assumed that the tidal debris could circularize and be viscously accreted by the SMBH on a timescale much shorter than $t_{\rm min}$, in which case the bolometric luminosity of the TDE would trace $\dot{M}_{\rm fall}$ \cite{Rees88} provided the radiative efficiency was time independent.  The details of disk formation and evolution needed to assess the validity of this assumption are sufficiently complicated that we shall defer them to \S \ref{sec:NewtonianAccretion}.

\subsection{Dynamics of Tidal Compression}
\label{sec:comp}

Although tidal stretching along the vector pointed towards the SMBH is responsible for tidal disruption, the disrupting star may also suffer severe compression in an orthogonal direction as it passes within the tidal radius $r_t$.  For the subset of TDEs with $\beta$ significantly greater than $1$, tidal free fall leads to a homologous vertical collapse of the star (vertical meaning orthogonal to the star's orbital plane).  The star pinches into an hourglass-like configuration as it passes through a caustic located near its orbital pericenter, but the buildup of hydrodynamic pressure causes its fluid elements to undergo a vertical bounce at this pinch point.

Again making use of the frozen-in approximation, we can study the tidal free fall (ballistic motion) of the disrupted star's matter inside the tidal sphere.  Importantly, in the $z$-direction orthogonal to the orbital plane, the tidal equations are $\Ddot{z}\propto -z$, implying a homologous vertical collapse.  The equations of tidal free fall can be solved exactly in Newtonian gravity \cite{Stone+13a} by using the parabolic Hill equations \cite{Sari+10}.  These solutions give an asymptotic vertical collapse speed (as fluid elements approach the caustic) of $v_c \simeq \beta (GM_\star/R_\star)^{1/2}$.  In the simplest version of the frozen-in approximation, the star will compress into an infinitely thin pancake at a pinch point shortly after pericenter passage.  In reality, complete stellar ``pancaking'' is prevented by two effects.  First, the star does not tidally disrupt all at once.  The leading edge begins collapsing before the trailing edge, and the collapsing star first assumes a pear-like shape, and then an hourglass-like one, as its fluid elements pass through the pinch point  \cite{Rosswog+09}.  Second, and more importantly, the star is not actually made of test particles, and severe compression at the vertical caustic will build up enough hydrodynamic pressure to lead to a rapid and violent hydrodynamical bounce at the pinch point.

Some of the earliest theoretical work on TDEs was motivated by this compression and bounce \cite{Carter&Luminet82}, which can increase the victim star's central density by several orders of magnitude for large values of $\beta$.  Tidal compression was first studied by means of the semi-analytic affine model \cite{Carter&Luminet83, Carter&Luminet85}, which is an implementation of the tensor virial theorem under the assumption that the disrupting star's geometry remains ellipsoidal\footnote{See Ref. \cite{Ivanov&Novikov01} for a more elaborate generalization of the affine model.}.  However, this process can also be understood by using simple analytic arguments.  Equating the kinetic energy of the vertical collapse $v_c^2 \simeq \beta^2 GM_\star/R_\star$ to the internal energy of a polytropic fluid with adiabatic index $\gamma$, we find that compression increases the average density of the star by a factor $\rho_c/\rho_\star \sim \beta^{2/(\gamma-1)}$, after which the hydrodynamic bounce begins.  Because this compression is primarily one-dimensional (the stellar cross-section within the orbital plane changes only by factors of order unity), the height of the star at peak compression is reduced by this same factor $z_c/R_\star \sim \beta^{-2/(\gamma-1)}$.  The temperature of the disrupting star is also dramatically enhanced by a factor $T_c/T_\star \sim \beta^2$.  In the limit of a uniformly ``pancaking'' star, the duration of peak compression is only $\tau_c \sim \tau_\star \beta^{-(\gamma+1)/(\gamma-1)}$, which, for large $\beta$, is much shorter than the star's dynamical time $\tau_\star = (G\rho_\star)^{-1/2}$.  These estimates imply that tidal compression is a very sensitive function of $\beta$: for a $\gamma=5/3$ polytrope, $\rho_c \propto \beta^3$ and $\tau_c \propto \beta^{-4}$.  A $\gamma=4/3$ polytrope exhibits even steeper dependencies, with $\rho_c \propto \beta^6$ and $\tau_c \propto \beta^{-7}$.

While the above scaling relations appear crude, they agree to order unity with the results of the semi-analytic affine model \cite{Carter&Luminet86}.  Similar levels of agreement are observed with the results of one-dimensional hydrodynamical simulations of stellar collapse (i.e. simulations of a collapsing column of star) for $\gamma=5/3$, although the agreement is not as good for the $\gamma=4/3$ case because of the much greater importance of shock dissipation \cite{Brassart&Luminet08}.  While large amounts of tidal compression are observed in fully three-dimensional simulations of stellar disruption, three-dimensional effects appear to reduce the total compression relative to the affine and one-dimensional predictions \cite{Rosswog+09, Guillochon+09}.  

The observational significance of tidal compression remains ambiguous.  The tidal squeezing of the star by the SMBH will enhance thermonuclear burning rates, but the total energy release is likely to be modest for main-sequence stars \cite{Guillochon+09}.  White dwarf TDEs are more promising sites for runaway thermonuclear burning \cite{Rosswog+09, 2017ApJ...839...81T, Tanikawa17a}, although these events will only occur frequently enough to be detectable if IMBHs are common in the universe (because of the lower Hills mass $M_{H} \lesssim 10^5 M_\odot$ for white dwarfs).  Tidal disruption of a white dwarf will likely ignite a large fraction of the star, and radioactive decay of the unbound debris will power an atypical Type Ia supernova that accompanies the accretion transient \cite{MacLeod+16}.  More speculatively, tidal compression may alter the gravitational-wave (GW) signal of the TDE in interesting ways, but we defer a discussion of this to \S \ref{sec:GRCompress}.  

Perhaps the most promising observable signature of tidal compression (at least for main-sequence stars) is a brief X-ray shock breakout as the shocks launched during peak compression expand outwards into the crushed star's photosphere \cite{Kobayashi+04}.  The shock breakout from the tidal disruption of a main-sequence star is expected to last $\sim 10-100~{\rm s}$ and may produce a peak X-ray luminosity $L \sim 10^{44}~{\rm erg/s}~(\beta/10)^{5/2}$ \cite{Guillochon+09}.  Such shock-breakout signals have not yet been detected, but could greatly aid the theoretical analysis of TDE light curves by pinpointing the moment of disruption.

\section{Accretion in Newtonian Gravity} \label{sec:NewtonianAccretion}

In the previous section, we saw that the broad hierarchy of energies intrinsic to the TDE problem allows us to understand the basics of tidal disruption from an analytic perspective which has been largely confirmed by hydrodynamic simulations.  However, the aftermath of tidal disruption is more complicated, as it depends on the subtle interplay of hydrodynamics, GR, and probably magnetohydrodynamic effects as well.  Modeling the competition between stellar self-gravity, black-hole gravity, and fluid dynamics accurately is difficult because of the disparate length scales and time scales of the system from disruption to accretion.  Analytic arguments will help us understand the evolution of tidal-debris streams and their assembly into an accretion flow, but hydrodynamic simulations are more important for studying these problems than they were in \S \ref{sec:basic}.  Existing TDE simulations generally use either Lagrangian smoothed particle hydrodynamics (SPH) techniques \cite{Bicknell&Gingold83,EvansKochanek89,Laguna+93,Lodato+09,Rosswog+09,Ayal+00,Bogdanovic04,Kobayashi+04,Hayasaki+13,Bonnerot+16,Hayasaki+16}, where the grid-free calculation reduces the computational expense with respect to length scale, or Eulerian, grid-based finite-volume methods \cite{Khokhlov1993a,Khokhlov1993b,Frolov+94,Diener+97,Guillochon+09,Guillochon&RamirezRuiz13,Cheng&Evans13,Cheng&Bogdanovic14} which follow the stellar debris locally instead of with respect to the black hole.  

In this section, we consider tidal-debris dynamics and accretion physics from a Newtonian perspective.  While this simplified picture may explain the late-time behavior of TDE accretion disks, it largely fails to explain the disk-formation process.  We will return to this problem in the context of relativistic gravity in \S \ref{sec:GR2}.

\subsection{Numerical Challenges}
\label{S:NumChal}

Modeling the physical processes involved in TDEs is challenging because they encompass a wide dynamic range.  One way to see this is through a comparison of characteristic length scales and time scales for the star and debris disk.  From Eq.~(\ref{eq:energy}), we derive a semi-major axis $a_{\rm min}=G M_\bullet/2\Delta\epsilon \simeq r_t^2/2R_\star$ for the most bound debris and $t_{\rm min}=2\pi(a_{\rm min}^3/GM_\bullet)^{1/2}$ for its orbital period.  Comparing these length and time scales to the stellar radius $R_\star$ and dynamical time $\tau_\star$ prior to disruption, we find
\begin{equation}\label{eq:dyn_length}
\frac{R_\star}{a_{\rm min}} \sim \left( \frac{M_\star}{M_\bullet} \right)^{2/3} \sim 10^{-4} \left (\frac{M_\bullet}{10^6 M_\odot} \right )^{-2/3} \left (\frac{M_\star}{M_\odot}\right )^{2/3},
\end{equation}
and
\begin{equation}\label{eq:dyn_time}
\frac{\tau_\star}{t_{\rm min}} \sim \left( \frac{M_\star}{M_\bullet} \right)^{1/2} \sim 10^{-3} \left ( \frac{M_\bullet}{10^6 M_\odot} \right )^{-1/2} \left ( \frac{M_\star}{M_\odot} \right )^{1/2}.
\end{equation}
A self-consistent hydrodynamic simulation of tidal disruption and accretion-disk formation would both resolve the star over fractions of its length scale $R_\star$ and contain the entire region over which the gaseous debris disk extends, which at least initially is a multiple of $a_{\rm min}$.  Furthermore, for the disk to form, the simulation must be evolved in time on the order of $t_{\rm min}$, which according to Eq.~(\ref{eq:dyn_time}) can be several orders of magnitude longer than the stellar dynamical time (note that $\tau_\star$ is comparable to the orbital period at the circularization radius $r_c$).  Since the requisite dynamic ranges in length and time increase with mass ratio, simulations of the aftermath of a TDE have often ``cheated'' by choosing artificial event parameters to mitigate this problem.  One such strategy is to consider TDEs caused by small IMBHs, typically with $M_\bullet \sim 10^3 M_\odot$ \citep{Rosswog+09, Haas+12, Guillochon+14, Shiokawa15, Evans+15}.  An alternative strategy is to use a SMBH of realistic mass, but to reduce the debris apocenter by considering a star consumed from an eccentric ($e \sim 0.9$) orbit rather than a nearly parabolic one with eccentricity given by Eq.~(\ref{E:eccstar}).  This will decrease $a_{\rm min}$ and $t_{\rm min}$, leading to a more tractable simulation \citep{Hayasaki+13, Bonnerot+16, Hayasaki+16, Sadowski+16}.  In general, simulating the disruption process itself is less computationally taxing than simulating the long-term evolution of the debris, and Lagrangian (smoothed particle hydrodynamics [SPH]) techniques have been used to simulate tidal disruption for decades \cite{Nolthenius&Katz82,Bicknell&Gingold83,EvansKochanek89}.  The primary numerical challenge arises from high-$\beta$ encounters, where tidal compression is severe and a high degree of mid-plane resolution (particle number for Lagrangian techniques, or grid cells for Eulerian ones) is required to capture the physics of the bounce discussed in \S \ref{sec:comp}.  As the modern simulations \cite{Mainetti+17, Guillochon&RamirezRuiz13} of low-$\beta$, Newtonian tidal disruption described in \S \ref{SS:NewtTDE} are relatively converged across computational techniques, we will now turn our attention to debris evolution.

\subsection{Accretion-Disk Formation}
\label{sec:NewtonianCircularization}

Stellar debris streams are born with very high eccentricities according to Eq.~(\ref{E:emin}), but some of the observed properties of TDE flares (e.g. thermal soft X-ray emission; see \S \ref{sec:obs}) are consistent with emission from a transient quasi-circular accretion disk.  How can dynamically cold debris streams circularize into such an accretion disk?  This question turns out to be difficult to answer in purely Newtonian gravity, as stellar debris on Keplerian orbits does not circularize into an accretion disk for realistic astrophysical parameters.  Current theoretical work suggests that general-relativistic phenomena are critical to the circularization process, as we will describe in \S \ref{sec:GR2}. In this section, we will briefly outline the phenomenology of tidal debris streams in Newtonian gravity.

In the immediate aftermath of a TDE, the most tightly bound debris has specific binding energy $\sim \Delta\epsilon \ll \epsilon_c$ according to Eq.~(\ref{E:Ec}).   This implies that an enormous amount of energy must be dissipated -- most likely in hydrodynamic shocks within or between debris streams -- if the debris is to circularize into a  circular accretion disk.  An early semi-analytic examination of tidal-stream dynamics \cite{Kochanek94} showed that debris streams are dynamically cold (i.e. pressure is unimportant), except at pericenter and apocenter, where vertical focusing of the orbits can create internal shocks within the stream.  Self-gravity is initially important for the debris streams\footnote{This is not true for deeply plunging encounters ($\beta \gtrsim 3$) where compression shocks inject so much heat into the disrupting star that the debris streams never recollapse \cite{Guillochon+14}.}, causing them to collapse into virial equilibrium (self-gravity balancing tidal shear) along their transverse dimensions while remaining axially unbound\footnote{Further fragmentation is possible for very stiff gas equations of state \cite{Coughlin+16}.}.

Hydrodynamical simulations of tidal disruption around IMBHs find significant debris circularization due to the shocks launched by stream compression at pericenter, but this is likely an artifact of the unrealistic mass ratio, as the fraction of orbital energy dissipated in the pericenter shock is $\sim \beta(M_\star/M_\bullet)^{2/3} \ll 1$ \cite{Guillochon+14}.  Simulations of stream evolution around SMBHs in Newtonian gravity find no significant dissipation at pericenter \cite{Hayasaki+13} or any other purely hydrodynamic mechanism to circularize the tidal debris.

\subsection{Accretion-Disk Evolution}
\label{sec:AccretionDiskEvolution}

While tidal-stream circularization appears unlikely in Newtonian gravity, GR provides hope that at least a large subset of TDEs will circularize their debris rapidly (\S \ref{sec:circularization}).  In this subsection, we will examine this subset of TDEs from the perspective of standard $\alpha$-disk theory, i.e. assuming that angular-momentum transport is mediated by a kinematic viscosity $\nu = \alpha c_s H$.  Here $c_s$ and $H$ are the radius-dependent disk sound speed and scale height, while $\alpha<1$ is the usual dimensionless Shakura-Sunyaev viscosity parameter \cite{Shakura&Sunyaev73}.  In reality, molecular viscosity is negligible in astrophysical plasmas, and angular-momentum transport in accretion disks is usually controlled instead by magnetohydrodynamic (MHD) turbulence \cite{Balbus&Hawley91, Hawley+95}.  We use this model to obtain a reasonable estimate of the emission properties without a full general relativistic magnetohydrodynamics (GRMHD) simulation.  If circularization is efficient, early-time mass return into the disk will occur at super-Eddington rates as indicated by Eq.~(\ref{eq:MDotNormed}), creating a geometrically thick accretion torus with $H/R \sim 1$.  In an $\alpha$-disk, the viscous inflow time at some radius $R$, with local azimuthal frequency $\Omega(R)$, is \begin{equation} \label{E:tvisc}
t_{\rm visc}=\alpha^{-1} \Omega^{-1}(R)\left(\frac{H}{R}\right)^{-2}.
\end{equation}
As $\Omega^{-1}(R) \ll t_{\rm min}$ (because $\Delta\epsilon \ll \epsilon_c$), Eq.~(\ref{E:tvisc}) implies that $t_{\rm visc} \ll t_{\rm min}$ unless $\alpha$ is extremely small.  The mass-inflow rate through the disk will therefore be $\dot{M}_{\rm acc} \approx \dot{M}_{\rm fall}$ \cite{Ulmer99}.  In other words, stellar debris accretes through the disk into the SMBH horizon shortly after returning to pericenter and circularizing into the disk via shocks.  Super-Eddington accretion flows are often approximated with analytic or semi-analytic ``slim disk'' models \cite{Abramowicz+88, Sadowski09}, and we can gain some insight with a simple slim-disk model designed to model the structure of quasi-stationary TDE disks at high Eddington ratio \cite{Strubbe&Quataert09}.  In this (Newtonian) model, the disk emits as a thermal, multi-color blackbody in which each radial annulus has an effective temperature
\begin{equation}
\sigma_{\rm SB} T_{\rm eff}^4 = \frac{3GM_\bullet \dot{M}_{\rm acc}f}{8\pi R^3}K^{-1},
\end{equation}
and an aspect ratio
\begin{equation}
\frac{H}{R}=\frac{3}{2}f\frac{10\dot{M}_{\rm acc}}{\dot{M}_{\rm Edd}} \frac{r_g}{R}K^{-1},
\end{equation}
where $\sigma_{\rm SB}$ is the Stefan-Boltzmann constant, $f=1-(R_{\rm in}/R)^{1/2}$ ($R_{\rm in}$ is the inner edge of the disk, which is less than or equal to the innermost stable circular orbit), and the dimensionless factor
\begin{equation}
K = \frac{1}{2} + \left[ \frac{1}{4}+6f\left(\frac{10\dot{M}_{\rm acc}}{\dot{M}_{\rm Edd}}\right)^2 \left(\frac{r_g}{R} \right)^2 \right]^{1/2}.
\end{equation}
The multi-color blackbody emission from such a disk, which extends from $R_{\rm in}$ to an outer radius $R_{\rm out} \simeq r_c$, will peak in the extreme UV or soft X-ray with bolometric luminosities $L \sim 10^{43-44}~{\rm erg~s}^{-1}$.  Making the (substantial) assumptions that $\dot{M}_{\rm acc} \propto \dot{M}_{\rm fall}$, and that the radiative efficiency $\eta$ is a constant in time, we expect the bolometric light curve to decline at late times as $L \propto t^{-5/3}$ as indicated by Eq.~(\ref{E:fallback}).  However, many TDEs are observed at optical wavelengths far down the Rayleigh-Jeans tail of emission from this compact slim-disk model.  At these wavelengths, the light curve $L_{\rm opt} \propto \dot{M}_{\rm acc}^{1/4} \propto t^{-5/12}$ \cite{Lodato&Rossi11}, and the peak luminosity $L_{\rm opt} \sim 10^{41}~{\rm erg~s}^{-1}$.

This simple, compact $\alpha$-disk model is in decent agreement with observed soft X-ray emission from a large population of TDE candidates (\S \ref{sec:obs}).  However, these predictions completely fail to explain the light curves or peak luminosities of optically bright TDE candidates.  Theoretical solutions to this ``optical excess'' generally fall into two categories.  The first category of solutions invokes an optically thick shroud of tidal debris from the disrupted star, which obscures a substantial solid angle on the sky and intercepts X-ray and extreme UV photons from the inner disk.  This reprocessing layer might be a quasi-static configuration of gas \cite{Loeb&Ulmer97, Guillochon+14, Coughlin&Begelman14, Roth+16} or an outflowing wind generated by the disk or the circularization process \cite{Metzger&Stone16, Roth&Kasen17}.  The second category of solutions argues that the observed optical emission is actually unrelated to accretion power, and is generated in the circularization process itself, as debris streams first dissipate, and then radiate, a fraction of their orbital-energy excess \cite{Lodato12, Piran+15, Shiokawa15}.  The true origin of optical emission in TDEs remains an open question at the time of writing.

In contrast to this compact, slim-disk model appropriate for super-Eddington accretion flows, the first models of TDE accretion disks were developed by solving the time-dependent partial differential equation governing radially spreading one-dimensional (azimuthally- and vertically-averaged) $\alpha$-disks \cite{Cannizzo+90}.  This early work focused on the very late-time evolution of the TDE disk, after the timescale hierarchy had reversed and $t_{\rm visc} \gg t_{\rm min}$.  The transition to this regime roughly occurs when the accretion rate becomes sub-Eddington, as $H/R$ rapidly declines once $\dot{M}_{\rm acc}(t) < \dot{M}_{\rm Edd}$ and $t_{\rm visc} \propto (H/R)^{-2}$.  In this regime, the accretion rate through the disk drops, as it is now regulated not by the addition of new mass but instead by the slow transport of angular momentum outward by internal stresses\footnote{However, at very late times, thermal instability may cause significant fluctuations in the accretion rate; see Ref. \cite{Shen&Matzner14} for more recent one-dimensional $\alpha$-disk modeling that accounts for limit-cycle instabilities in a viscously spreading TDE disk.}.  The TDE accretion disk spreads outwards well beyond $r_c$ so as to conserve angular momentum, and its bolometric luminosity declines as $L \propto t^{-1.2}$.

We have simplified this analysis by neglecting three important physical differences between transient TDE disks and the better-studied steady-state accretion disks of most other accreting astrophysical black holes:
\begin{enumerate}
    \item TDE disks are generically inclined with respect to the equatorial plane (perpendicular to the spin) of a Kerr SMBH \cite{Stone&Loeb12}. Because stars approach the SMBH isotropically in the Newtonian limit, TDE disks probably form with substantial inclination angles.  As the effects of this inclination are fundamentally general relativistic, we will return to them in \S \ref{sec:diskPrecession}.
    \item TDE disks may be poorly magnetized.  Stars carry little net magnetic flux with them as they approach the SMBH (in comparison to the total magnetic flux in steady-state AGN disks) and may only acquire significant net magnetic flux under special circumstances \cite{Kelley+14}.  Some GRMHD simulations \cite{Sadowski+16} suggest that this reduces the efficiency of the magneto-rotational instability (MRI) responsible for angular-momentum transport (and thus the effective $\alpha$ viscosity compared to typical accretion disks).  If true, accretion through a TDE disk may only be possible through exotic alternative angular-momentum transport mechanisms \cite{Svirski+17, Nealon+18}.
    \item TDE disks will be highly eccentric if circularization is inefficient \cite{Guillochon+14, Shiokawa15}.  The evolution of highly eccentric accretion disks is not a solved problem and likely leads to qualitative changes in important accretion phenomena, such as the MRI \cite{Chan+17}.
\end{enumerate}

In summary, significant questions remain regarding: (i) bulk hydrodynamic evolution of TDE accretion disks, assuming they form in the first place, and (ii) the origin of optical emission in observed TDE candidates.  Both of these topics seem to be intimately linked to our earlier questions (\S \ref{sec:NewtonianCircularization}) concerning the circularization of debris, which do not appear answerable in Newtonian gravity.  We shall therefore return in greater detail to these subjects when discussing general-relativistic treatments of stellar tidal debris in \S \ref{sec:GR2}.

\section{Tidal Disruption Event Rates in Newtonian Gravity} \label{sec:rates}

TDEs occur when stars find themselves on orbits with pericenter $r_p \lesssim r_t$, but how often does this happen in real galaxies?  Early work by Hills conjectured that the TDE rate could be high enough to explain the growth of SMBHs and observed AGN activity \cite{Hills75}.  This optimistic estimate of the TDE rate was predicated on the assumption that stellar relaxation (due to two-body scattering) was efficient enough to rapidly replenish regions of orbital phase space that would be depleted by tidal disruption over an orbital period.  Such orbits are said to lie within the ``loss cone'' of the SMBH, since at a point a distance $r$ from the SMBH, the velocity vectors of stars on doomed orbits ($r_p \le r_t$) occupy a cone with axis directed towards the SMBH and opening angle $\theta \simeq (r_t/r)^{1/2}$.  Loss cone orbits have specific angular momentum $L \le L_t$, and, as mentioned before, a specific energy that is almost always $|\epsilon| \ll GM_\bullet / r_t$. 

It was quickly realized, however, that that the diffusion of stars into the loss cone is inefficient within a critical radius $r_{\rm crit}$, defined as the point where the specific angular momentum diffusion coefficient, $\langle \Delta L^2 \rangle$ equals $L_t^2$ \cite{1976MNRAS.176..633F, 1977ApJ...211..244L}.  For $r<r_{\rm crit}$, the phase space loss cone is empty and the rate of TDEs is set by the rate at which relaxational processes repopulate it.  Generally, most TDEs originate from radii $r\sim r_{\rm crit}$.  These early analytic estimates were confirmed by a full numerical solution to the two-dimensional Fokker-Planck equation governing the diffusion of stars in energy and angular momentum space \cite{1978ApJ...226.1087C}.

The accurate calculation of TDE rates, and the recognition that inefficient star formation could leave large reservoirs of gas at galactic centers, undermined the conjecture that TDEs are the primary feeding mechanism responsible for SMBH growth and AGN activity.  However, early X-ray detections of potential TDE candidates \cite{1995Natur.378...39R, Bade96, KomossaBade99} and the discovery of tight correlations between SMBH masses and properties of their host galaxies \cite{1995ARA&A..33..581K, 1998AJ....115.2285M, 2000ApJ...539L...9F, 2000ApJ...539L..13G} renewed interest in predicting TDE rates \cite{1999MNRAS.309..447M}.  The TDE rate in a spherically symmetric galactic nucleus depends sensitively on both the SMBH mass $M_\bullet$ and the logarithmic slope $\Gamma$ of its surface brightness profile at small radii \cite{2004ApJ...600..149W}.  Smaller SMBHs ($M_\bullet \lesssim 10^{8} M_\odot$) are typically hosted by ``power-law'' galaxies ($\Gamma > 0.5$), while more massive SMBHs ($M_\bullet \gtrsim 10^{8.5} M_\odot$) generally live in ``core'' galaxies ($\Gamma < 0.3$).  The TDE rate per galaxy can be roughly parametrized as
\begin{equation}
\dot{N} = \dot{N}_0 \left( \frac{M_\bullet}{10^8\, M_\odot} \right)^{-B}
\end{equation}
with $B > 0$; galaxies with smaller SMBHs have higher TDE rates because (i) empirically, their stellar distributions are more steeply cusped (higher $\Gamma$) and (ii) smaller SMBHs have smaller critical radii $r_{\rm crit}$ at which there are higher stellar densities \cite{2004ApJ...600..149W}.  Using the observed $M_\bullet-\sigma$ correlation between SMBH mass and host-galaxy velocity dispersion available at the time, Ref. \cite{2004ApJ...600..149W} predicted $\dot{N}_0 = 2.1 \times 10^{-4}~{\rm yr}^{-1}$ and $B = -0.25$ for galaxies with a stellar density distribution given by the singular isothermal sphere, $\rho(r) = \sigma^2/2\pi Gr^2$.  A more recent estimate by Ref. \cite{Stone&Metzger16} using a broad sample of 144 galaxies  \cite{2007ApJ...662..808L, 2007ApJ...664..226L} predicted $\dot{N}_0 = 2.9 \times 10^{-5}~{\rm yr}^{-1}$ and $B = -0.404$ for the entire galaxy sample. Observations of the black-hole mass function of optical TDEs \citep{2018ApJ...852...72V} are consistent with these predictions for $B$; today's number of detected TDEs is small, but a very steep dependence of the flare rate on SMBH mass, $|B|>1$, can be ruled-out.  As smaller SMBHs dominate the total TDE rate, the observation (or lack thereof) of TDEs in dwarf galaxies is a powerful probe of the existence of intermediate-mass black holes in such systems \cite{2004ApJ...600..149W}.

In addition to the total TDE rate, we are also interested in the distribution of penetration factors $\beta$, as this parameter can affect observed TDE properties.  The dimensionless diffusivity parameter $q(\epsilon)\equiv \langle \Delta L^2(\epsilon) \rangle / L_t^2$ can be calculated as a function of SMBH mass $M_\bullet$ and stellar density profile; portions of phase space with $q \gg 1$ are rapidly repopulated by stellar diffusion (the full loss-cone or ``pinhole'' limit), while those with with $q \ll 1$ remain unoccupied (the empty loss-cone or ``diffusive'' limit) \cite{1977ApJ...211..244L}.  Defining the pinhole fraction $f_{\rm pinhole}$ as the fraction of TDEs originating in portions of phase space with $q > 1$, the TDE rate per unit penetration factor can be approximated 
as \cite{Kochanek16}
\begin{equation}
\frac{d\dot{N}}{d\beta} \approx 
\begin{cases}
f_{\rm pinhole}\beta^{-2}(1-\beta_{\rm max}^{-1}) \quad \quad &\beta \gtrsim 1  \\
 1 - f_{\rm pinhole} \quad \quad &\beta \approx 1,
 \end{cases}
\end{equation}
where $\beta_{\rm max} \simeq (L_t/L_{\rm cap})^2$, and the angular momentum threshold for direct capture by the event horizon of a non-spinning SMBH is $L_{\rm cap} = 4M_\bullet$.  The pinhole fraction scales with SMBH mass as
\begin{equation}
f_{\rm pinhole} = 0.22 \left( \frac{M_\bullet}{10^8\, M_\odot} \right)^{-0.307},
\end{equation}
with a TDE rate-weighted value $f_{\rm pinhole} \approx 0.3$ over the entire galaxy sample \cite{Stone&Metzger16}.  This suggests that $\beta \geq 1$ events will be reasonably common in the entire TDE sample, but $\beta \simeq 1$ will dominate for the more massive SMBHs $M_\bullet \gtrsim M_{\rm GR}$ for which relativistic effects will be important.  A substantial caveat to this section is that we have treated disruption in a Newtonian way.  As we will show in \S \ref{sec:GR1}, our Newtonian threshold for disruption ($L_t$) will be altered by general-relativistic effects when $M_\bullet \gtrsim M_{GR}$, and our threshold for capture ($L_{\rm cap}$) will acquire a dependence on SMBH spin and orbital inclination.

We should also note that the calculations presented in this section assumed that two-body scattering was responsible for the stellar diffusion that refills the loss cone.  Perturbations that break the spherical symmetry of the stellar density profile can generate torques that more rapidly refill the loss cone through resonant relaxation \cite{1996NewA....1..149R, 2013degn.book.....M}, as can axisymmetry \cite{1999MNRAS.309..447M, 2013ApJ...774...87V} or triaxiality \cite{2004ApJ...606..788M} in the stellar density profile.

\section{Tidal Disruption in General Relativity}
\label{sec:GR1}

GR modifies the dynamical evolution of the victim star, its tidal debris streams, and the accretion disk that may be produced following tidal disruption.  In this section, we focus on how GR affects the tides that disrupt the victim star and the resulting orbits of the tidal debris, while in \S \ref{sec:GR2} we examine the effects of relativistic precession on the circularization of tidal streams into an accretion disk and the subsequent evolution of this disk.

\subsection{Relativistic Tides}

In Newtonian gravity, an object like a star experiences tides when there is a spatial gradient in the gravitational force acting on it.  Such a spatial gradient will cause two nearby particles in this object, initially with the same velocity, to experience different accelerations and change their relative positions unless the object's internal forces are strong enough to preserve its integrity.  In GR, gravity affects the motion of particles by determining the metric of spacetime, which in turn determines the geodesics on which freely falling particles move.  The separation between two nearby particles of an object with timelike 4-velocity $u^\mu$ can be specified by a spacelike deviation 4-vector $X^\beta$ which evolves with proper time $\tau$ along the geodesic according to the geodesic-deviation equation
\begin{equation} \label{E:geoD}
\frac{d^2X^\beta}{d\tau^2} = u^\mu \nabla_\mu (u^\alpha \nabla_\alpha X^\beta) = -R^\beta_{~\mu\alpha\nu} u^\mu X^\alpha u^\nu = -C^\beta_{~\alpha} X^\alpha .
\end{equation}
Here $\nabla_\mu$ are covariant derivatives, $d/d\tau \equiv u^\alpha \nabla_\alpha$ is the derivative with respect to the
proper time, $R^{\beta}_{~\mu \alpha \nu}$ is the Riemann curvature tensor, and
\begin{equation} \label{E:tidaltensor}
C^{\beta}_{~\alpha} \equiv R^{\beta}_{~\mu \alpha \nu}u^{\mu}u^{\nu}
\end{equation}
is the tidal tensor.  The self-gravity of a non-relativistic object like a main-sequence star can be treated as an internal force independent of the external tidal field; in GR, such a star will avoid tidal disruption if the internal acceleration due to self-gravity exceeds the tidal acceleration specified by the geodesic-deviation equation.

The geodesic-deviation equation (\ref{E:geoD}) is coordinate-independent, with Greek spacetime indices $\alpha, \beta$ that run from 0 to 3.  We can solve this equation more easily by choosing to use local Fermi normal coordinates ($\tau, X^{(i)}$) valid in a neighborhood of spacetime about the center of mass of the star \cite{Manasse&Misner+63, 1983grg1.conf..438L}.  These coordinates define a tetrad of orthonormal 4-vectors: the star's 4-velocity $\lambda^\mu_{~(0)} \equiv u^\mu$ and three spacelike 4-vectors $\lambda^\mu_{~(i)}$ chosen to be parallel transported along the geodesic.  By projecting the deviation 4-vector $X^\alpha$ and tidal tensor $C^\beta_{~\alpha}$ onto this tetrad:
\begin{subequations} \label{E:FNC}
\begin{align}
X^{(i)} &= X^\alpha \lambda_\alpha^{~(i)} \\ \label{E:TTFNC}
C^{(i)}_{\quad(j)} &= C^\beta_{~\alpha} \lambda_\beta^{~(i)} \lambda^\alpha_{~(j)} = R^\beta_{~\mu\alpha\nu} \lambda_\beta^{~(i)} \lambda^\mu_{~(0)} \lambda^\alpha_{~(j)} \lambda^\nu_{~(0)}~,
\end{align}
\end{subequations}
the geodesic-deviation equation becomes
\begin{equation} \label{E:geoDFNC} 
\frac{d^2X^{(i)}}{d\tau^2} = -C^{(i)}_{\quad(j)} X^{(j)}.
\end{equation}
where the Latin indices $i, j$ now only run from 1 to 3.  The tidal tensor $C^{(i)}_{\quad(j)}$ is a symmetric $3 \times 3$ matrix with three real eigenvalues and eigenvectors.  One of these eigenvalues will be negative, implying that the object will be stretched along the direction of the corresponding eigenvector.  A star traveling along a geodesic of the SMBH's spacetime will be tidally disrupted if this stretching, determined by the negative eigenvalue of the tidal tensor, exceeds the star's self-gravity.

Eq.~(\ref{E:geoDFNC}) describes the quadrupole tide that dominates for infinitesimal deviations $X^{(j)}$, but it receives corrections from the octopole and higher-order tides \cite{Manasse&Misner+63, Mashhoon+75, Marck+83, Ishii+05, Cheng&Evans13}
\begin{equation} \label{E:geoDFNCHO} 
\frac{d^2X^{(i)}}{d\tau^2} = -C^{(i)}_{\quad(j)} X^{(j)} - \frac{1}{3} C^{(i)}_{\quad(j)(k)} X^{(j)} X^{(k)} + \ldots
\end{equation}
where
\begin{equation}
C^{(i)}_{\quad(j)(k)} = (\nabla_\delta R^\beta_{~\mu\alpha\nu}) \lambda_\beta^{~(i)} \lambda^\mu_{~(0)} \lambda^\alpha_{~(j)} \lambda^\nu_{~(0)}\lambda^\delta_{~(k)}~.
\end{equation}
These higher-order tides are suppressed at the tidal radius by a factor $R_\star/r_t = (M_\star/M_\bullet)^{1/3}$, indicating that they are sub-percent level corrections for tidal disruptions by SMBHs but can be of order $\sim 10\%$ for IMBHs.

To understand relativistic tides around spinning SMBHs, we must briefly review the geodesics of the Kerr metric \cite{1963PhRvL..11..237K}.  In Boyer-Lindquist coordinates \cite{1967JMP.....8..265B} and units where $G = c = 1$, the Kerr metric takes the form
\begin{eqnarray} \label{E:met}
ds^2 &=& -\left( 1 - \frac{2M_\bullet r}{\Sigma} \right) dt^2
- \frac{4M_\bullet ar \sin^2 \theta}{\Sigma} dt d\phi
+ \frac{\Sigma}{\Delta} dr^2
\nonumber \\ 
&&  + \Sigma d\theta^2 + \left( r^2 + a^2 +
\frac{2M_\bullet a^2r \sin^2 \theta}{\Sigma} \right) \sin^2 \theta d\phi^2~,
\end{eqnarray}
where $a = \chi_\bullet M_\bullet$ for SMBHs with dimensionless spin magnitude $\chi_\bullet$, $\Sigma \equiv r^2 + a^2 \cos^2 \theta$, and $\Delta \equiv r^2 - 2Mr + a^2$.  This metric possesses a timelike Killing field $\xi^\mu$ and an axial Killing field $\psi^\mu$ \cite{1984ucp..book.....W}, implying that a massive particle with 4-velocity $u^\mu$ will have conserved specific energy $E \equiv -g_{\mu\nu} u^\mu \xi^\nu$ and angular momentum $L_z \equiv g_{\mu\nu} u^\mu \psi^\nu$.  Note that this relativistic definition of the specific energy includes the particle's rest mass, and this definition of the angular momentum corresponds to the component of the orbital angular momentum parallel to the SMBH spin in the Newtonian limit.  The Kerr metric also possesses a Killing tensor $K_{\mu\nu}$ \cite{1970CMaPh..18..265W} which can be used to define the conserved Carter constant $Q$ \cite{1968PhRv..174.1559C}:
\begin{equation}
Q \equiv K_{\mu\nu} u^\mu u^\nu - (L_z - aE)^2~.
\end{equation}
In the Newtonian limit, $Q \to L_\perp^2$, where $L_\perp$ is the magnitude of the component of the specific orbital angular momentum perpendicular to the SMBH spin.  This suggests that we can define a conserved angular momentum magnitude $L \equiv (Q + L_z^2)^{1/2}$ and inclination $\cos\iota \equiv L_z/L$.  For the parabolic geodesics on which tidally disrupted stars travel, we can also define an argument of pericenter $\omega$ and longitude of ascending node $\Omega$ far from the SMBH; the five constants $(E, L, \iota, \omega, \Omega)$ fully specify a Kerr geodesic except for a trivial time translation.  The axisymmetry of the Kerr metric implies that geodesics with different values of $\Omega$ are related by a trivial rotation about the SMBH spin axis.

As the tidal tensor $C^{(i)}_{\quad(j)}$ given by Eq.~(\ref{E:TTFNC}) depends on the 4-velocity $\lambda^\mu_{~(0)} = u^\mu$ along a Kerr geodesic, the criterion for tidal disruption in GR is not just $r_p < r_t$ (or equivalently $L < L_t$) as in Newtonian gravity, but whether the geodesic lies within a three-dimensional hypersurface of the four-dimensional parameter space $(E, L, \iota, \omega)$ of physically distinct Kerr geodesics.  This surface depends on both the SMBH mass $M$ and spin magnitude $\chi_\bullet$.  As most tidally disrupted stars are on nearly parabolic orbits, only the intersection of this surface with the plane $E = 1$ is physically relevant for TDEs.  Furthermore, the peak tidal acceleration on Kerr geodesics only depends on the argument of pericenter $\omega$ at the sub-percent level even for highly inclined, low angular momentum orbits around highly spinning Kerr SMBHs \cite{ServinNotes}.  Ignoring this mild $\omega$ dependence, tidal disruption will occur in GR for $L < L_d(a, \iota)$ or equivalently $\beta > \beta_d(a, \iota)$.  This function will be largely independent of the inclination $\iota$ for SMBH masses $M_\bullet \ll M_{GR}$, where tidal disruption occurs at $r \gg r_t$, but will develop a significant $\iota$ dependence for $M_\bullet \gtrsim M_{GR}$ and high spins $a \simeq M_\bullet$.  For a fixed angular momentum $L$, relativistic tides are stronger on retrograde ($\iota = \pi$) than prograde ($\iota = 0$) orbits.

GR also effects the orbits of the tidal streams produced following a TDE.  In the ``frozen-in'' approximation in Newtonian gravity, a particle located at $X^{(i)}$ with respect to the center of mass of a star on a parabolic orbit will have a specific binding energy
\begin{equation} \label{E:EXNewt}
E(\mathbf{X}) = -X^{(i)} [\partial_{(i)} \Phi]_{r_t} = -\mathbf{X} \cdot \mathbf{\hat{r}} \frac{GM_\bullet}{r_t^2}~.
\end{equation}
Here the gradient of the gravitational potential $\partial_{(i)}\Phi$ is evaluated at the tidal radius $r_t$ where the tidal force exerted by the SMBH balances the star's self-gravity.  The zeroth-order terms in $X^{(i)}$ vanish because the star is initially on a parabolic orbit ($E = 0$), while the gradient of the kinetic energy vanishes because all particles in the unperturbed star are assumed to be moving at the same velocity at the time of disruption in the ``frozen-in'' approximation.  This expression is maximized for $\mathbf{X} = -R_\star \mathbf{\hat{r}}$, the particle on the star's surface closest to the SMBH, and has a maximum value of $\Delta \epsilon$ in agreement with Eq.~(\ref{eq:energy}).  In GR, this same expression (Eq. \ref{E:EXNewt}) holds, except that the ordinary flat-space gradient must be replaced by a covariant derivative and the Newtonian potential must be replaced by the expression $E \equiv -g_{\mu\nu} \lambda^\mu_{~(0)} \xi^\nu$ for the relativistic specific energy \cite{2012PhRvD..86f4026K}:
\begin{equation} \label{E:delErel}
E(\mathbf{X}) = X^{(i)} \lambda_{~(i)}^\alpha \nabla_\alpha \left[ -g_{\beta\gamma} \lambda_{~(0)}^\beta \xi^\gamma \right] = -g_{\beta\gamma} \lambda_{~(0)}^\beta X^{(i)} \lambda_{~(i)}^\alpha \Gamma_{\alpha t}^\gamma~.
\end{equation}
In the last equality, we have used the vanishing covariant derivative of the metric, the assumption of the ``frozen-in'' approximation that all the particles have the same 4-velocity $\lambda_{~(0)}^\beta$ as the center of mass, and the fact that the covaient derivative of the timelike Killing field $\xi^\gamma$ is given by the indicated Christoffel symbols.  In the ``frozen-in'' approximation, the covariant derivative in Eq.~(\ref{E:delErel}) should be evaluated at the point along the star's geodesic where the tidal acceleration given by the geodesic-deviation equation (\ref{E:geoDFNC}) equals the star's self gravity.  Similar expressions to Eq.~(\ref{E:delErel}) exist for the angular momentum $L_z$ and Carter constant $Q$ of the tidal debris, but these results only provide first-order corrections, and to zeroth-order the tidal debris inherits the the values of $L_z$ and $Q$ of its stellar progenitor.

Eqs.~(\ref{E:EXNewt}) and (\ref{E:delErel}) determine the energy distribution $dM/dE$ of the tidal debris in Newtonian gravity and GR respectively, and thus the fallback accretion rate $\dot{M}_{\rm fall}$ onto the SMBH according to Eq.~(\ref{E:fallback}).  As the tidal debris spends most of its time near apocenter far from the SMBH, one can use the Newtonian expression for $dE/dt$ even in the relativistic case.  In principle, one can compare these two equations to predict how the fallback accretion rate $\dot{M}_{\rm fall}$ differs in Newtonian gravity and GR, but this comparison must be performed carefully, as it is not obvious which Keplerian orbit should be compared to a given relativistic geodesic \cite{2017PhRvD..95h3001S}.  A gauge-invariant choice, as we are considering parabolic orbits that extend to large distances from the SMBH at which the two theories of gravity converge, is to compare a geodesic with angular momentum magnitude $L = (Q + L_x^2)^{1/2}$ to the Keplerian orbit with the same angular momentum.  For non-spinning SMBHs described by the Schwarzschild metric, this comparison shows that stars experience stronger tides at pericenter in GR than they do in Newtonian gravity on orbits with the same angular momentum $L$ \cite{2017PhRvD..95h3001S}.  This implies that the penetration factor threshold $\beta_d$ for full tidal disruption is lower in GR, and that stars only modestly above this threshold may only experience partial disruption in Newtonian gravity.  Although one might expect that the stronger tides in GR lead to more tightly bound tidal debris and a higher fallback accretion rate, if one evaluates the gradients appearing in Eqs.~(\ref{E:EXNewt}) and (\ref{E:delErel}) at the points on each orbit experiencing the same threshold for tidal disruption (a gauge-invariant criterion), one finds that the tidal debris is slightly {\it less} tightly bound in GR.  For the most massive Schwarzschild SMBH capable of fully disrupting a star without capture by the event horizon (for which relativistic effects will be most significant), the most tightly bound debris is $\sim 77\%$ as tightly bound in GR as it is in Newtonian gravity \cite{2017PhRvD..95h3001S}.  If the TDE bolometric luminosity traced the fallback accretion rate as is sometimes assumed, this would imply slightly fainter TDEs in GR compared to Newtonian gravity.

\begin{figure}
\begin{center}
\includegraphics[scale=1.4]{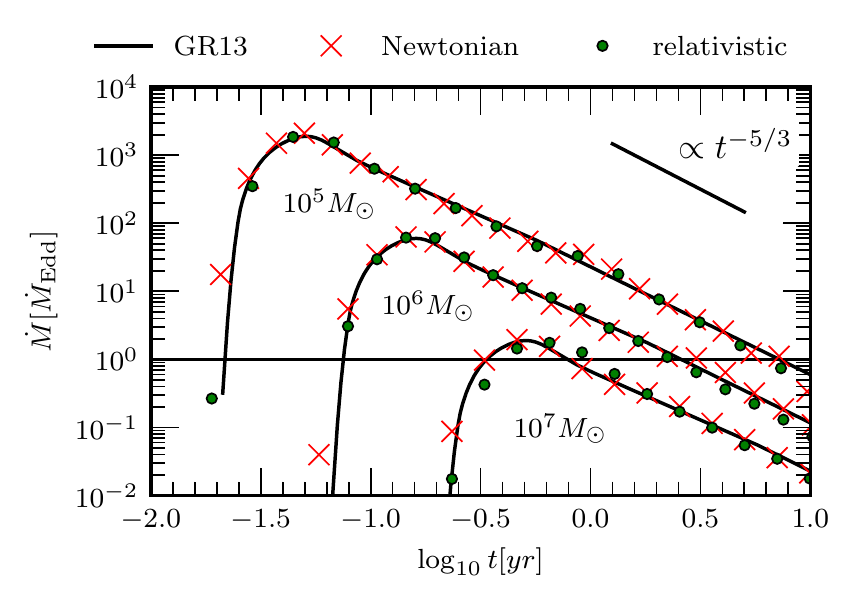} 
\end{center}
\caption{Newtonian and relativistic predictions for the fallback accretion rate $\dot{M}$ onto the SMBH following a TDE, normalized by the Eddington rate $\dot{M}_{\rm Edd}$ of Eq.~(\ref{eq:edd}), as a function of the time $t$ following disruption.  Results are given for TDEs of main-sequence stars by non-spinning SMBHs of masses $10^5 M_\odot$, $10^6 M_\odot$, and $10^7 M_\odot$.  The predictions of the Newtonian and relativistic simulations of Cheng and Bogdanovi\'{c}~\cite{Cheng&Bogdanovic14} are labeled red Xs and green circles respectively.  The black curves are numerical fits to the Newtonian simulations of Guillochon and Ramirez-Ruiz \cite{Guillochon&RamirezRuiz13}.  At late times, $\dot{M} \propto t^{-5/3}$ as predicted by Eq.~(\ref{E:fallback}).  This figure was originally published as the left panel of Fig.~7 of Ref.~\cite{Cheng&Bogdanovic14}.}  \label{figure:return_rate}
\end{figure}

\begin{figure}
{ \begin{center}
\includegraphics[scale=1.3]{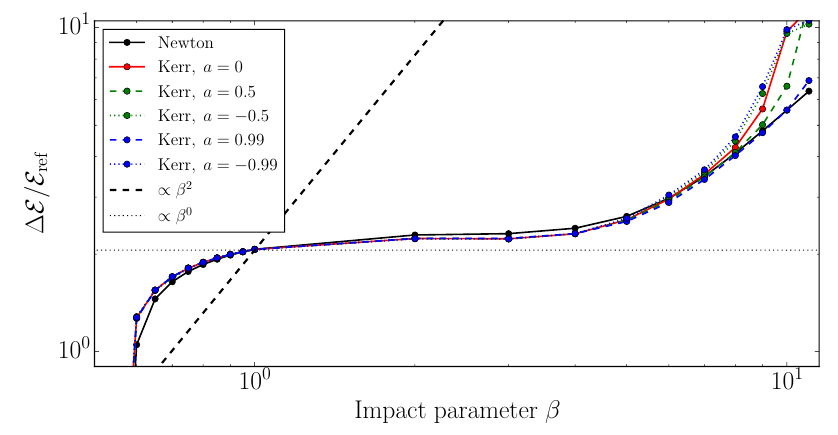}
\end{center}
\caption{The spread in specific binding energy [normalized by $\Delta\epsilon$ given by Eq. (\ref{eq:energy})] of the tidal debris produced in simulations of the disruption of a Solar-type star ($M_\star = M_\odot, R_\star = R_\odot$) by a $10^6 M_\odot$ SMBH as a function of penetration factor $\beta$.  The solid black curve shows simulations using Newtonian gravity, while the solid red curve shows TDEs by a non-spinning SMBH in GR.  The dashed (dotted) green curves correspond to TDEs of stars on prograde (retrograde) orbits about Kerr SMBHs with $a = 0.5 M_\bullet$, while the corresponding blue curves show TDEs by a SMBH with $a = 0.99 M_\bullet$.  Good agreement is seen with the ``freezing model'' until very high $\beta$ values, where an enhancement to the energy spread may arise from a combination of the hydrodynamic bounce and general-relativistic effects.  This figure was originally published as Fig.~12 of Ref. \cite{Tejeda+17}.}  \label{fig:Tejeda} } 
\end{figure}

This estimate of relativistic effects using the ``frozen-in'' approximation can be tested using hydrodynamic simulations performed in a local frame following the star \cite{Frolov+94,Diener+97,Ishii+05,Cheng&Evans13}.  Using such a frame, Cheng and Bogdanovi\'{c} \cite{Cheng&Bogdanovic14} simulated TDEs by non-spinning SMBHs in Newtonian and relativistic gravity and compared the predicted energy distributions of the tidal debris and corresponding fallback accretion rates.  These simulations were performed from $20(r_p^3/GM)^{1/2}$ before to $40(r_p^3/GM)^{1/2}$ after pericenter passage, capturing the entire tidal disruption process, and included both octupole and hexadecapole tides.  The predicted fallback accretion rates for TDEs of Solar-type stars ($M_\star = M_\odot, R_\star = R_\odot$) by SMBHs with masses $M_\bullet = 10^5 M_\odot$, $10^6 M_\odot$, and $10^7 M_\odot$ are shown in Fig.~\ref{figure:return_rate}.  We see that GR has a modest effect on the fallback accretion rate even for $M_\bullet = 10^7 M_\odot$ where it is expected to be the most significant.  For this SMBH mass, GR reduces the peak fallback accretion rate and increases the time that it remains super-Eddington by $\sim 15\%$, consistent with the predictions of Ref.~\cite{2017PhRvD..95h3001S} that the stronger tides in GR lead to disruption at larger radii and less tightly bound tidal debris.  The comparisons shown in Fig.~\ref{figure:return_rate} were made between Newtonian and relativistic simulations with the same gauge-dependent pericenter $r_p$ but different angular momenta $L$; the robustness of the predicted fallback accretion rates validates the use of the ``frozen-in'' approximation at the instant of tidal disruption.  The global SPH simulations of TDEs by spinning SMBHs in Tejeda {\it et al.} \cite{Tejeda+17} also support the ``frozen-in'' approximation by showing that both penetration factor $\beta$ and SMBH spin have little effect on the debris energy distribution and fallback accretion rates except for highly penetrating encounters ($\beta \gtrsim 8$) where relativistic precession or severe tidal compression may be significant; their results are illustrated in Fig \ref{fig:Tejeda}.  Greater departures from Newtonian predictions were also seen in the simulation in Ref.~\cite{Cheng&Bogdanovic14} of the full disruption of a white dwarf on an orbit with substantial apsidal precession ($L \simeq 4.08 M$ only marginally above the capture threshold $L_{\rm cap} = 4M$ at which precession diverges).  We discuss capture by the event horizon in greater detail in the following section, and elaborate on the effects of relativistic precession in \S \ref{sec:GR2}.

\subsection{Capture by the Event Horizon} \label{S:cap}

One feature of SMBHs in GR entirely lacking in Newtonian gravity is the existence of event horizons, boundaries defining regions of spacetime from which light cannot escape to future null infinity.  Stars or tidal debris traveling on Kerr geodesics with $L < L_{\rm cap}(a, \iota)$ will be directly captured by the SMBH event horizon, and thus be unable to produce a luminous TDE.  One can determine $L_{\rm cap}(a, \iota)$ by finding the values of $L_z$ and $Q$ (or equivalently $L$ and $\iota$) for which the right-hand side of the equation of motion:
\begin{equation}
\Sigma^2 \left( \frac{dr}{d\tau} \right)^2 = R(r) = [E(r^2 + a^2) - aL_z]^2 - \Delta[r^2 + (L_z - aE)^2 + Q]
\end{equation}
has a double root \cite{1976PhRvD..14.3281Y}.  For parabolic orbits ($E = 1$), this simplifies to
\begin{equation}
R(r) = 2M_\bullet r(r^2 + a^2 - 2aL\cos\iota + L^2) - L^2[r^2 + a^2(1 - \cos^2\iota)]~.
\end{equation}
For Schwarzschild SMBHs ($a = 0$), $L_{\rm cap} = 4M_\bullet$, while for Kerr SMBHs ($a \neq 0$), $L_{\rm cap}$ acquires a dependence on $\iota$ which for maximally spinning ($a = M_\bullet$) SMBHs ranges from $L_{\rm cap} = 2M_\bullet$ on prograde ($\iota = 0$) orbits to $L_{\rm cap} = 2(1 + \sqrt{2})M_\bullet$ on retrograde ($\iota = \pi$) orbits.  Observable TDEs will be produced by stars on geodesics that lead to disruption but not capture, i.e. geodesics with specific angular momentum $L$ and inclination $\iota$ in the range $L_{\rm cap}(a, \iota) < L < L_d(a, \iota)$.  To lowest order in $M_\star/M_\bullet \ll 1$, the threshold for disruption $L_d \propto M_\bullet^{2/3}$ like $L_t$ defined following Eq.~(\ref{E:beta}).  The threshold for capture scales as $L_{\rm cap} \propto M_\bullet$ since GR is scale invariant.  This implies that SMBHs above the relativistic Hills mass $M_{\rm H}(a, \iota)$ for which $L_d = L_{\rm cap}$ will be incapable of producing observable TDEs.

\begin{figure}
{ \begin{center}
\includegraphics[scale=0.5]{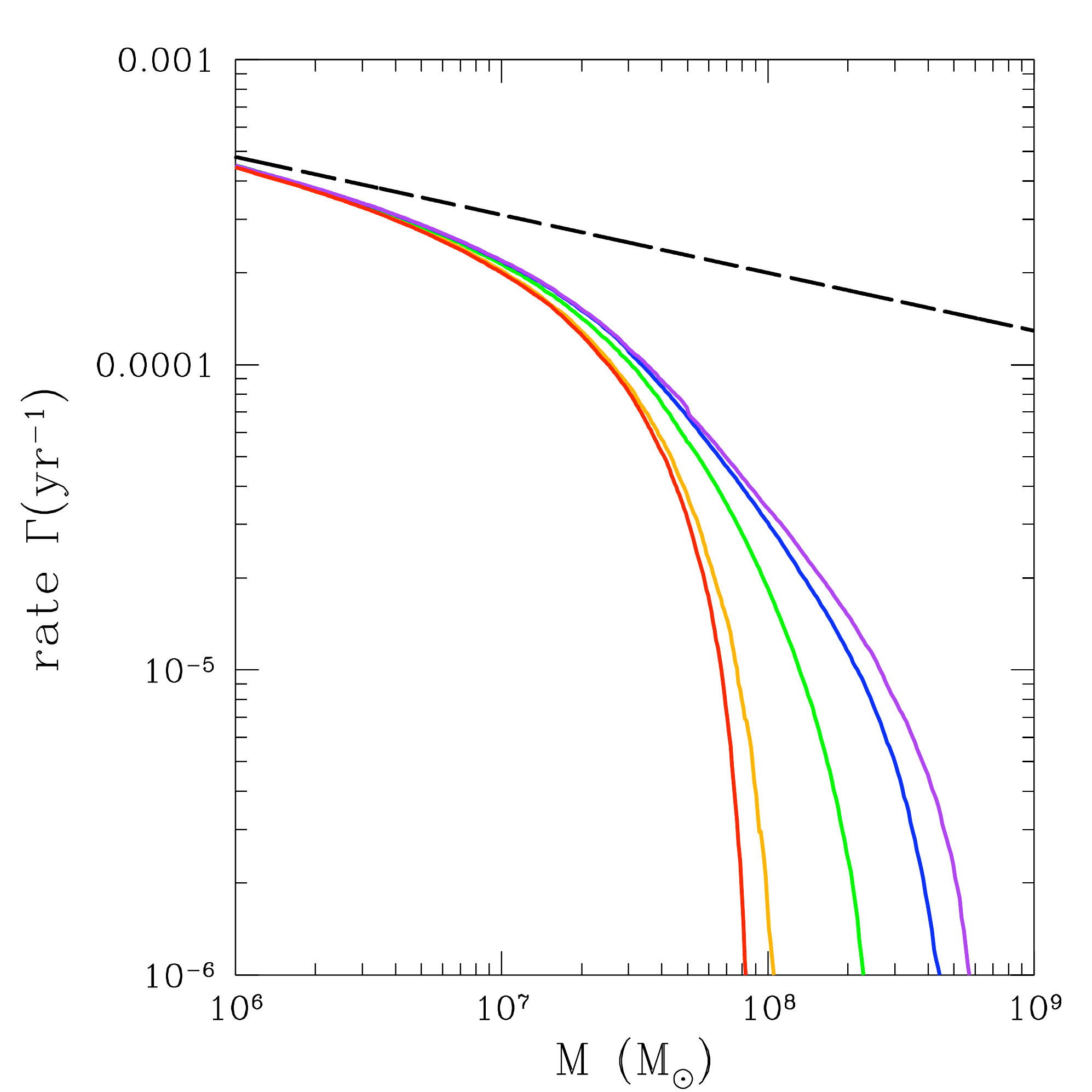}
\end{center}
\caption{The TDE rate per year $\Gamma$ in galaxies with a singular isothermal sphere stellar density profile $\rho = \sigma^2/2\pi Gr^2$ as a function of SMBH mass $M$.  The dashed black line is the Newtonian prediction \cite{2004ApJ...600..149W}, while the colored curves show the suppression due to capture by the event horizon for SMBHs with spins $\chi_\bullet = 0$ (red), 0.5 (orange), 0.9 (green), 0.99 (blue), and 0.999 (purple).  This figure previously appeared as Fig.~4 in Kesden \cite{2012PhRvD..85b4037K}.}  \label{F:supp} } 
\end{figure}

For Schwarzschild SMBHs, stars with solar mass and radius will fail to disrupt above $M_{\rm H}(0, \iota) \simeq 10^{7.4} M_\odot$, a factor $\sim 4.5$ larger than would have been predicted if the stronger tides in GR were not taken into account \cite{2017PhRvD..95h3001S}.  For Kerr SMBHs, $M_{\rm H}$ depends on both the spin magnitude $a$ and the orbital inclination $\iota$ \cite{Beloborodov+92}.  Although tides are stronger on higher inclination orbits of the same specific angular momentum $L$, the capture threshold $L_{\rm cap}$ has an even stronger inclination dependence implying that $M_{\rm H}(a, \iota)$ will be a monotonically decreasing function of inclination $\iota$ \cite{2006A&A...448..843I,2012PhRvD..85b4037K}.  In Fig.~\ref{F:supp}, we show that as $M_\bullet$ increases above $M_{\rm H}(a, \iota = \pi)$, the TDE rate by Kerr SMBHs will become highly suppressed by direct capture, with complete suppression (for all inclinations) occurring for $M_\bullet > M_{\rm H}(a, \iota = 0)$.  This upper limit $M_{\rm H}(a, \iota = 0)$ can be as large as $\sim 10^9 M_\odot$ for Solar-type stars \cite{1994MmSAI..65.1135S, 2006A&A...448..843I}.

Direct capture will eliminate emission from TDEs with $\beta > \beta_{\rm cap}$ by SMBHs of all masses, but this will be a mild effect for $M_\bullet \ll M_{\rm H}$ as the rate of high $\beta$ encounters scales as $\beta^{-1}$ in the full loss-cone limit and is exponentially suppressed\footnote{Although see Ref. \cite{Weissbein&Sari17}.} in the empty loss-cone regime \cite{Stone&Metzger16}.  For small SMBHs in the full loss-cone regime with $M_\bullet < M_{\rm H}(a, \iota = \pi)$, the suppression of the rate of observable TDEs is almost entirely independent of SMBH spin magnitude $a$, because the reduced number of stars captured from orbits with inclinations $\iota < \pi/2$ is almost entirely compensated by an increased number of stars captured from orbits with inclinations $\iota > \pi/2$ \cite{1977ApJ...212..367Y}.  Preliminary results in the empty loss-cone regime relevant for massive SMBHs where relativistic effects are significant suggest that Kerr SMBHs will preferentially disrupt stars on retrograde orbits for $M_\bullet \lesssim M_{\rm H}(a, \iota = \pi)$, with this preference shifting towards prograde orbits for $M_\bullet \lesssim M_{\rm H}(a, \iota = 0)$.

Fig. \ref{F:supp} demonstrates how spin-dependent relativistic tides and direct capture imprint themselves into the statistical distribution of TDE host masses.  The super-exponential suppression seen above $M_\bullet \sim M_{\rm H}(a, \iota = \pi)$ is a purely relativistic prediction, absent from Newtonian gravity, and one with a nontrivial $a_\bullet$-dependence.  With a large enough sample of observed TDEs and well characterized SMBH masses, this general-relativistic prediction will be observationally testable, and indeed, existing data has offered preliminary confirmation of it, as we will see in \S \ref{sec:ObsGR}.

\subsection{Relativistic Tidal Compression}
\label{sec:GRCompress}

The process of Newtonian tidal compression described in \S \ref{sec:comp} can be heavily modified by the stronger tidal fields present in GR.  Relativistic tidal compression will also differ from the Newtonian case because of the different center-of-mass trajectories of parabolic ($E = 1$) orbits in GR.  The effects of GR on tidal compression were first studied by combining the semi-analytic affine model with the tidal tensor $C^{(i)}_{~(j)}$ expressed in Fermi normal coordinates \cite{Luminet&Marck85}.  The primary result of this modeling was to show that multiple caustics may exist in the congruence of geodesics along which tidal debris travels following tidal disruption.  In other words, orbits with $L \lesssim 3L_{\rm cap}$ will commonly experience two phases of compression and bounce\footnote{Three or more vertical collapses are possible in a small portion of parameter space.}.  Using geometric arguments, one can derive an approximate criterion for the existence of a double caustic: the center-of-mass trajectory (comparable to that of the victim star) must self-intersect inside the tidal radius \cite{Luminet&Marck85}.  Even weakly relativistic TDEs that only possess a single vertical caustic will see modified collapse behavior compared to Newtonian gravity; the stronger GR tidal field leads to an earlier vertical collapse.  The predictions of the affine model for relativistic collapse are in good agreement with one-dimensional hydrodynamical simulations \cite{Brassart&Luminet10}, but have not been extensively tested against three-dimensional simulations.  There has also been little investigation of relativistic collapse in the Kerr metric, although this may be particularly interesting for inclined orbits.  The spherical symmetry of the tidal tensor for non-spinning SMBHs implies that vertical collapse should be homologous (at least in the test-particle limit), but the changing directions of the eigenvectors of the tidal tensor along inclined geodesics of the Kerr metric may lead to significantly non-homologous collapse \cite{Stone+13a}.

One interesting but speculative outcome of tidal compression in GR is possible GW emission.  A weak, low-frequency GW signal is expected to arise from the changing orbital quadrupole moment of the star-SMBH system, with characteristic frequencies and strain amplitudes \cite{Kobayashi+04, East14}
\begin{align}
    f_{\rm GW} \sim & \left(\frac{GM_\bullet}{r_p^3}\right)^{1/2} \sim 6\times10^{-4}~{\rm Hz} ~\beta^{3/2} \left( \frac{M_\star}{M_\odot} \right)^{1/2}\left( \frac{R_\star}{R_\odot} \right)^{-3/2}, \\
    h_{\rm GW} \sim & \frac{GM_\star r_g}{c^2 D r_p} \sim 2\times 10^{-22}~\beta \left( \frac{M_\bullet}{10^6 M_\odot} \right)^{2/3}\left( \frac{M_\star}{M_\odot} \right)^{4/3}\left( \frac{R_\star}{R_\odot} \right)^{-1} \left( \frac{D}{10~{\rm Mpc}} \right)^{-1}, \notag
\end{align}
where $D$ is the distance to the source.  This characteristic frequency $f_{\rm GW}$ is accessible to space-based laser interferometers like {\it LISA}, but the predicted amplitude $h_{\rm GW}$ is so weak that this signal would not be visible for TDEs outside the Local Group.  However, the changing quadrupole moment of the star itself will produce a second kind of GW signal \cite{Guillochon+09}.  Because the internal quadrupole moment of the star can vary much faster than the star's dynamical time $\tau_\star$, tidal compression may serve as a type of frequency amplifier that transfers some power into high-frequency GWs with $f \sim \tau_c^{-1} \sim \tau_\star^{-1} \beta^{(\gamma + 1)/(\gamma - 1)}$, where $\gamma$ is the adiabatic index as in \S \ref{sec:comp}.  This analytic argument suggests that main-sequence TDEs may radiate GWs with $f>10~{\rm Hz}$ for $\beta \gtrsim 15$, but the expected wave strain is very weak ($h \lesssim 10^{-24}$) \cite{Stone+13a}.  White-dwarf disruptions could radiate at $f>10~{\rm Hz}$ for $\beta \gtrsim 4$, but such high-frequency GWs have not been seen in numerical-relativity simulations of high-$\beta$ tidal disruption \cite{Rosswog+09, Haas+12}.  This may indicate that three-dimensional effects (specifically, the desynchronization of stellar collapse away from the pancaking limit) shift the GW power back to low frequencies.

\section{General Relativistic Accretion}
\label{sec:GR2}

In \S \ref{sec:GR1}, we presented well understood applications of GR to the tidal disruption process.  Now that we have exhausted these comparatively straightforward aspects of relativistic tidal disruption, we shall explore some more uncertain - but perhaps even more important - ways in which the outcome of a TDE can be affected by GR.  First, we will focus on how general-relativistic precession appears to control the circularization and disk-assembly process in the aftermath of a TDE.  Second, we will consider how SMBH spin may lead to precession of a compact TDE accretion disk, assuming one has been able to circularize in the first place.  We emphasize that both of these topics are, at the time of publication, very open questions among TDE researchers.  The solutions to these questions are likely to be nontrivial, but each holds significant promise for probing SMBH spin with TDEs.

\subsection{General Relativistic Circularization}
\label{sec:circularization}

To understand the early stages of accretion following a TDE, we must address the circularization problem defined in \S \ref{sec:NewtonianCircularization}: detailed, three-dimensional hydrodynamical simulations fail to efficiently circularize tidal debris in Newtonian gravity.  However, even very early analytic works on TDEs recognized the centrality of general-relativistic effects for circularization and the disk-formation process.  Apsidal precession forces highly eccentric debris streams to self-intersect, creating hydrodynamic shocks that dissipate large amounts of orbital energy and drive the circularization process \cite{Rees88}.  Conversely, circularization may be hindered by relativistic nodal precession, which arises primarily from Lense-Thirring frame dragging \cite{Cannizzo+90}.  Nodal precession winds debris streams into different orbital planes, and for some combinations of spin-orbit misalignment, spin magnitude, and stream thermodynamics, the failure of stream centerlines to directly intersect can strongly delay the disk-formation process \cite{Kochanek94}.

The debris streams will self-intersect at a radius dictated by the apsidal shift per radial period, which at lowest PN order is
\begin{equation} \label{E:apshift}
\Delta\omega_{\rm 1PN} = 3\pi\beta \left( \frac{r_g}{r_t} \right) = 0.2\beta \left( \frac{R_\star}{R_\odot} \right)^{-1} \left( \frac{M_\star}{M_\odot} \right)^{1/3} \left( \frac{M_\bullet}{10^6\,M_\odot} \right)^{2/3},
\end{equation}
implying a self-intersection radius \cite{Dai15}
\begin{align}
r_{\rm si} &= \frac{(1+e_{\rm min}) r_t}{\beta[1-e_{\rm min} \cos(\Delta\omega_{\rm 1PN} /2)]} \label{eq:rsi} \\
&\simeq 100 r_t \left( \frac{M_\star}{M_\odot} \right)^{-1/3} \left( \frac{M_\bullet}{10^6\,M_\odot} \right)^{1/3} \qquad \qquad \qquad (M_\bullet \lesssim 10^6 M_\odot,\, \beta \simeq 1), \nonumber \\
&\simeq \frac{18.6 r_t}{ \beta^3} \left( \frac{R_\star}{R_\odot} \right)^2 \left( \frac{M_\star}{M_\odot} \right)^{-2/3} \left( \frac{M_\bullet}{10^7\,M_\odot} \right)^{-4/3} \quad (M_\bullet \gtrsim 10^7 M_\odot,\, R_\star \lesssim R_\odot). \nonumber
\end{align}
We see from Eq.~(\ref{eq:rsi}) that debris self-intersection is a sensitive function of penetration factor $\beta$.  High-$\beta$ disruptions around small SMBHs, or any disruptions around large SMBHs, will favor self-intersections at small radii where orbital energy can be efficiently dissipated in shocks.  Less relativistic TDEs may be more common, however, as they are produced by the abundant $\beta \sim 1$ disruptions around small SMBHs that dominate the volumetric event rate.  In these TDEs, the natural stream self-intersection is far from the SMBH, with $r_{\rm si} \simeq a_{\rm min}$ for the most tightly bound debris.  Circularization will likely proceed much less efficiently in this regime.  For example, the observed X-ray brightening of the TDE candidate ASASSN-15oi one year after its discovery in the UV/optical may have been delayed due to inefficient circularization around the $\sim 10^6 M_\odot$ SMBH \cite{2017ApJ...851L..47G}.

General-relativistic effects were first incorporated into hydrodynamical simulations of tidal disruption in the PN SPH framework of Ref.~\cite{Laguna+93}.  While these simulations were not run long enough to witness long-term stream evolution, they did capture important relativistic predictions (e.g. a $\beta=10$ run exhibited the double compression and bounce \cite{Luminet&Marck85} expected for deeply penetrating encounters).  A similar PN SPH approach was later used to model long-term stream evolution around non-spinning SMBHs \cite{Ayal+00}.  This pioneering work found significant accretion of gas following pericenter return.  Contrary to subsequent hydrodynamical simulations, it also found that a majority of initially bound debris is unbound by shock heating\footnote{This discrepancy may be due to the low particle number used in these SPH simulations.}.  

The issue of general-relativistic stream evolution was revisited by Ref.~\cite{Hayasaki+13}, which used a SPH code with a pseudo-Newtonian potential tailored to accurately reproduce the apsidal shift for high-$e$ orbits around a Schwarzschild SMBH \cite{Wegg12}.  This work was the first to simulate disk formation around a SMBH due to stream self-intersections, albeit with an artificially low initial stellar eccentricity ($e=0.8$).  It found rapid circularization through self-intersection shocks in simulations using the pseudo-Newtonian potential, but essentially zero circularization of debris when a purely Newtonian potential was used instead.  Similar SPH simulations ($M_\bullet \sim 10^6 M_\odot$, $1-e \sim 0.1$) were performed with other pseudo-Newtonian potentials \cite{Tejeda&Rosswog13} while varying gas thermodynamic properties \cite{Bonnerot+16}, and in post-Newtonian potentials up to 2.5PN order \cite{Hayasaki+16}.  In all of these approaches, reasonably efficient debris circularization was observed, in contrast to some Eulerian general-relativistic hydrodynamics simulations which found the formation of a highly elliptical accretion flow and little debris circularization \cite{Shiokawa15}.  This may not be altogether surprising, as the simulations in Ref. \cite{Shiokawa15} focused on the disruption of a white dwarf ($M_\star = 0.64 M_\odot$) by a small IMBH ($M_\bullet =500M_\odot$) with $\beta = 1$.  For this choice of parameters, Eq.~(\ref{eq:rsi}) predicts a self-intersection radius $r_{\rm si} \simeq 8.2 r_t$ comparable to $a_{\rm min} \simeq 4.6 r_t$, for which dissipation will be minimally effective.

\begin{figure}
{ \begin{center}
\includegraphics[scale=1.3]{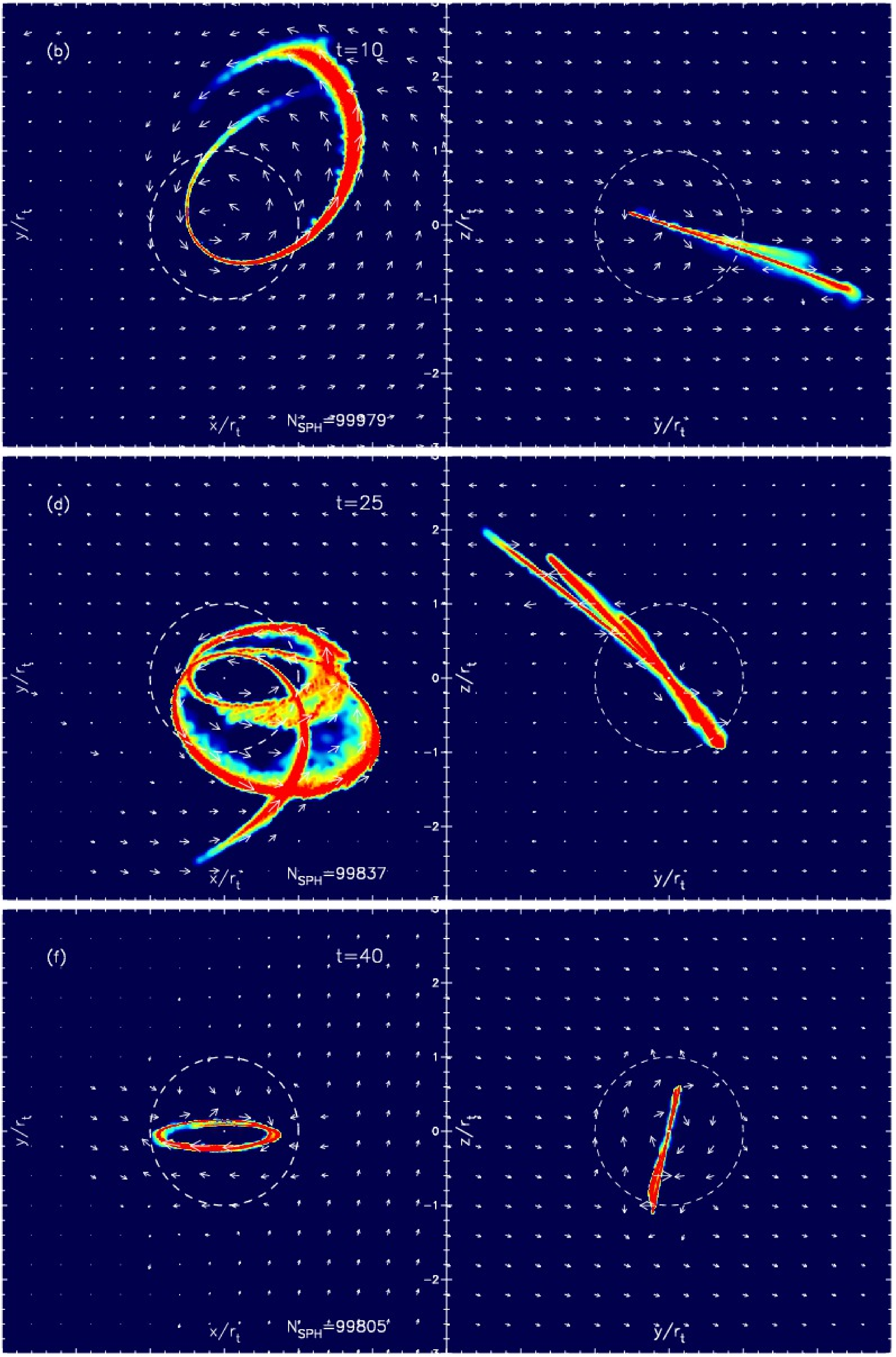} 
\end{center}
\caption{\label{figure:circ_retrograde} 
Projected surface density of the tidal stream produced following the tidal disruption of a Solar-type star ($M_\star = M_\odot, R_\star = R_\odot, \gamma = 5/3$) by a SMBH with $M_\bullet = 10^6 M_\odot$.  The SMBH has dimensionless spin $\chi_\bullet = 0.9$ and the star is initially on an orbit with semi-major axis $a_\star = 5r_t/3$, eccentricity $e_\star = 0.7$, and inclination $\iota = 90^\circ$.  The left column shows projections into the $xy$ plane perpendicular to the initial stellar angular momentum ($\mathbf{\hat{L}} = \mathbf{\hat{z}}$), while the right column shows projections into the $yz$ plane perpendicular to the SMBH spin ($\mathbf{\hat{S}} = -\mathbf{\hat{x}}$).  The top, middle, and bottom rows show snapshots at times $t = 10$, 25, and 40 after the start of the simulation, expressed in units of the period of a circular orbit at the tidal radius.  This figure was originally published as part of Fig.~16 of Hayasaki {\it et al.} \cite{Hayasaki+16}.
}  }
\end{figure}

Recent post-Newtonian SPH simulations \cite{Hayasaki+16} have captured debris circularization around Kerr SMBHs via inclusion of 1.5PN spin-orbit coupling terms.  For equatorial orbits, the circularization process is not too altered by the inclusion of SMBH spin.  The 1.5PN term for apsidal precession
\begin{align}
\Delta\omega_{\rm 1.5PN} &= -3\pi\sqrt{2}\chi_\bullet\beta^{3/2} \cos\iota \left( \frac{r_g}{r_t} \right)^{3/2} \\
&= -0.0409\chi_\bullet\beta^{3/2}\cos\iota  \left( \frac{R_\star}{R_\odot} \right)^{-3/2} \left( \frac{M_\star}{M_\odot} \right)^{1/2} \left( \frac{M_\bullet}{10^6\,M_\odot} \right) \nonumber
\end{align}
implies that for retrograde orbits ($\cos\iota <0 $), the total apsidal precession $\Delta\omega = \Delta\omega_{\rm 1PN} + \Delta\omega_{\rm 1.5PN}$ is enhanced, and the self-intersection radius $r_{\rm si}$ is correspondingly reduced according to Eq.~(\ref{eq:rsi}).  This leads to greater energy dissipation at the intersection point and more rapid circularization.  Prograde spin, ($\cos\iota >0$) conversely, delays circularization, but these differences are modest since typically $\Delta\omega_{\rm 1PN} \gg |\Delta\omega_{\rm 1.5PN}|$.  These simulations also confirmed the long-standing hypothesis that misaligned orbits around Kerr SMBHs may experience dramatic circularization delays because nodal precession causes the tidal stream to depart from the victim star's initial orbital plane, thus delaying self-intersection.  The lowest-order nodal shift (per radial orbit) is due to Lense-Thirring frame dragging, and is
\begin{align}
\Delta\Omega_{\rm 1.5PN} &= \pi\sqrt{2}\chi_\bullet\beta^{3/2} \left( \frac{r_g}{r_t} \right)^{3/2} \\
&= 0.0136\chi_\bullet\beta^{3/2} \left( \frac{R_\star}{R_\odot} \right)^{-3/2} \left( \frac{M_\star}{M_\odot} \right)^{1/2} \left( \frac{M_\bullet}{10^6\,M_\odot} \right). \nonumber
\end{align}
For very relativistic pericenters, the lowest-order nodal shift $\Delta\Omega_{\rm 1.5PN}$ becomes comparable to higher-order terms, such as the next-order, 2PN contribution from the SMBH quadrupole moment \cite{Merritt+10}.  In principle, the comparison of simulations of circularization that included these higher-order terms to observations of early-time TDE light curves could serve as an interesting probe of the no-hair theorem (as the quadrupole moment should be determined completely by $M_\bullet$ and $\chi_\bullet$).  In practice, this type of test may be hampered by the competing effects of gas physics.

Specifically, circularization delays depend on the details of tidal-stream thermodynamics.  Debris streams in SPH simulations that use an adiabatic (no-cooling) equation of state quickly puff up due to heating from internal shocks, and their cross-section becomes so large that no realistic amount of nodal precession can prevent self-intersection.  In the opposite thermodynamic limit (isothermal equation of state; fast cooling), the debris streams remain narrow enough that the circularization time can be $\gtrsim 10 t_{\rm min}$, with orbital energy only dissipating very slowly, through a succession of weakly grazing stream self-intersections as seen in Fig.~\ref{figure:circ_retrograde}.  

Much recent progress on the relativistic circularization problem has also come from semi-analytic models, which may hold significant promise for the future.  One such model used exact Kerr geodesic solutions to estimate the time until the onset of stream self-intersections \cite{Guillochon&RamirezRuiz15}, accounting for both apsidal and nodal precession; another such model used post-Newtonian orbital approximations to estimate the rate of circularization after the first self-intersections \cite{Bonnerot+17}.  While these models make a number of simplifying assumptions, they efficiently cover the parameter space of a very numerically intractable problem, and are deserving of further development.

Although the circularization problem is not yet solved, disk formation in TDEs appears to be a rare astrophysical arena where GR provides not merely a PN correction, but instead is the zeroth-order determinant of a directly observable outcome.  More surprisingly, the same can be said (albeit with less confidence) about SMBH spin and frame-dragging effects, which in most astrophysical environments represent even higher-order PN corrections beyond those expected for a Schwarzschild SMBH.  It is easy to understand, however, why relativistic effects play such a fundamental role in TDE disk formation.  The self-intersection shocks required to form an actual accretion disk are completely absent in Newtonian gravity, where the apsidal shift $\Delta\omega$ vanishes exactly.  Likewise, in the Schwarzschild metric, the nodal shift $\Delta\Omega$ vanishes exactly; it is only in the Kerr metric (and only for misaligned orbits) where accretion can be substantially delayed by misaligned SMBH spin.  

While the complex details of TDE circularization must imprint themselves on the early, rising portions of a TDE flare's light curve, models that describe the circularization process from first principles have yet to be developed.  At the time of writing, it seems plausible that the misaligned spin component of the SMBH will have a leading-order impact on this early phase of a TDE light curve, but this prediction must ultimately be confirmed by increasingly sophisticated hydrodynamical (and perhaps magnetohydrodynamical) simulations.

\subsection{Global Precession of TDE Disks}
\label{sec:diskPrecession}

One way in which TDEs are astrophysically unique is that they are generically {\it tilted} accretion systems.  In most other contexts, accretion disks around astrophysical black holes are expected to align with the black hole equatorial plane due to differential nodal precession caused by frame-dragging \cite{1975ApJ...195L..65B}.  However, as short-lived, transient systems, TDE disks may not have time to align themselves with the SMBH.  Furthermore, tidally disrupted stars approach the SMBH with a quasi-isotropic angular distribution\footnote{If the star cluster and potential surrounding the SMBH are spherical, the distribution of angular momenta $\mathbf{L}$ of tidally disrupted stars will be isotropic (flat in $\cos\iota$, $\omega$, and $\Omega$) in Newtonian gravity.  If these quantities are non-spherical, they may imprint anisotropies in the distribution of $\mathbf{L}$.  These anisotropies will reflect features of the potential on the scale of the influence radius $r_h \sim {\rm pc}$, and are thus unlikely to favor or disfavor equatorial orbits unless the SMBH and nuclear star cluster share a common origin.}, and are thus expected to produce disks with generically nontrivial inclination angle $\psi$.  In this subsection, we will ignore the complicated details of circularization discussed in \S \ref{sec:circularization}.  Given that circularization is an open problem, and one that may depend in a nontrivial way on Lense-Thirring precession, it is quite plausible that the circularization process biases TDE disk tilt angles $\psi$ away from the asymptotic orbital inclinations $\iota$ of the progenitor stars.  Here we will work under the simplifying assumptions that (i) a subset of TDEs are able to circularize their debris rapidly, and (ii) the orbital planes of stellar debris are not universally aligned with the SMBH equatorial plane during the circularization process.  Future simulations of TDE circularization will be needed to check the validity of these assumptions.

Let us assume a compact, tilted accretion disk: what happens next?  There is an extensive analytic and semi-analytic literature describing the evolution of tilted disks which transport angular momentum through the $\alpha$-viscosity approximation \cite{Shakura&Sunyaev73}.  It is important to note that this viscosity prescription only approximately captures physical angular-momentum transport (which, in real accretion disks, is mediated by turbulent magnetic stresses), and may have particular problems describing the behavior of {\it tilted} disks.  In the theory of tilted $\alpha$-disks, there are two different limits of behavior.  For thin disks, with $\alpha>H/R$, warps produced by differential nodal precession propagate diffusively \cite{Pringle92}, and alignment of the inner regions is expected.  However, for the thick disks characteristic of TDEs ($\alpha < H/R$, at least initially), warps propagate as bending waves \cite{Papaloizou&Lin95}, and the disk can precess as a quasi-rigid body.

We will follow Ref. \cite{Fragile+07} to analytically approximate the global precession rate of an inclined disk.  To simplify the calculation, we assume a simple disk structure: its 2D surface density profile $\Sigma(R) \propto R^{-\zeta}$, with an inner edge at $R_i$ and an outer edge at $R_o$.  Neglecting alterations to this profile from internal twist or bending waves, we first compute the total disk angular momentum:
\begin{equation}
J = 2\pi \int^{R_o}_{R_i} \Sigma(R) \Omega(R)R^3 dR,
\end{equation}
and next compute the integrated Lense-Thirring torque on all disk annuli:
\begin{equation}
\mathcal{N}= \frac{4\pi\chi_\bullet(GM_\bullet)^2}{c^3} \sin\psi \int^{R_o}_{R_i} \Sigma(R) \Omega(R) dR.
\end{equation}
Here $\Omega(R) = (GM_\bullet/R^3)^{1/2}$ is the azimuthal frequency of a circular orbit of radius $R$, and $\psi$ is the tilt angle between the disk and the SMBH equatorial plane.  Making the further approximation of rigid-body precession, we can write the precession time analytically as
\begin{equation}
T_{\rm prec} = \frac{2\pi J}{\mathcal{N}\sin\psi} = \frac{\pi r_g(1+2\zeta)}{c\chi_\bullet (5-2\zeta)} \frac{r_o^{5/2-\zeta} r_i^{1/2+\zeta}(1-(r_i/r_o)^{5/2-\zeta})}{1-(r_i/r_o)^{1/2+\zeta}}, \label{eq:tPrec}
\end{equation}
where we have normalized $r_o \equiv R_o/r_g$ and $r_i \equiv R_i/r_g$.  This estimate of the rigid-body precession timescale is crude, as it assumes Newtonian gravity, and a very simple disk structure.  This simple disk structure is in reasonable agreement with GRMHD simulations of tilted, thick accretion flows \cite{Fragile+07}.   

\begin{figure}
{ \begin{center}
\includegraphics[scale=1.00]{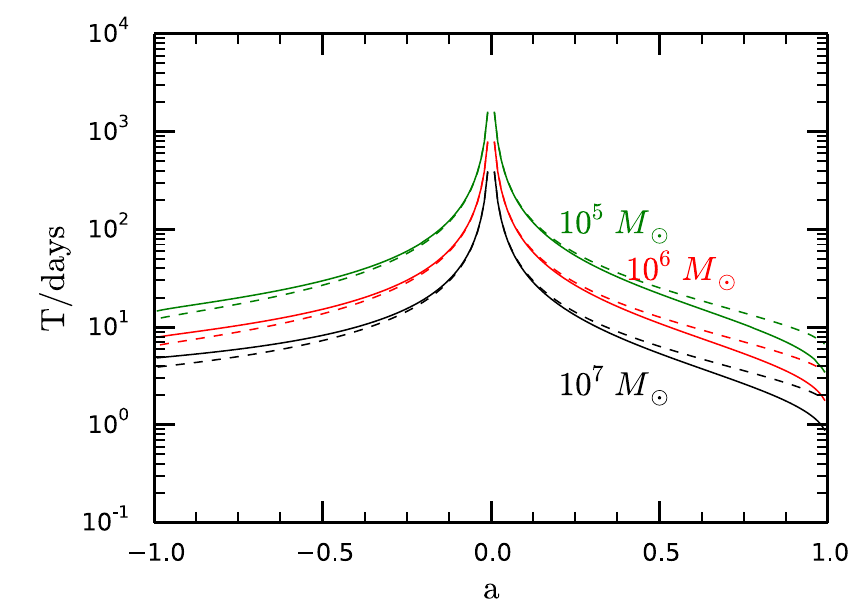}
\end{center}
\caption{\label{figure:PrecessionRate}
The global disk precession time $T$ as a function of dimensionless SMBH spin $a$.  Results for SMBHs with masses $M_\bullet = 10^5 M_\odot$, $10^6 M_\odot$, and $10^7 M_\odot$ are shown by the green, red, and black curves respectively.  This figure was originally published as Fig.~4 of Franchini {\it et al.} \cite{Franchini+16}.
}  }
\end{figure}

Franchini {\it et al.} \cite{Franchini+16} used the simple, analytic slim disk model presented in \S \ref{sec:NewtonianAccretion} to calculate the precession time $T_{\rm prec}$ given by Eq.~(\ref{eq:tPrec}) as a function of the dimensionless SMBH spin $\chi_\bullet$ as shown in Fig.~\ref{figure:PrecessionRate}.  If $\chi_\bullet \gtrsim 0.2$, the global precession time of TDE disks will be $\sim 1-30$ days, a plausibly observable range.  This estimate neglects the gradual increase in disk precession periods that will occur if the disk spreads outwards viscously \cite{Shen&Matzner14}.  If this viscous spreading occurs, $T_{\rm prec}$ will grow monotonically and may eventually quench the precession of the TDE disk.  Precession may likewise turn off once $\dot{M} \ll \dot{M}_{\rm Edd}$ and the disk becomes geometrically thin, leading to alignment through the Bardeen-Petterson effect \cite{Stone&Loeb12}, or after viscous processes internal to a globally precessing disk force alignment \cite{Franchini+16}.

The universality of global precession in tilted, thick accretion disks is an open question.  Nearly rigid-body precession of geometrically thick disks is a clear prediction of $\alpha$-disk theory and is also generally seen in purely hydrodynamical simulations that use this simplified viscosity prescription \cite{Larwood+96}.  However, it is unclear whether angular-momentum transport through magnetized turbulence should yield the same behavior.  Early Eulerian GRMHD simulations of mildly tilted ($\psi = 15^\circ$), thick accretion flows found rigid-body precession at rates in reasonable agreement with Eq.~(\ref{eq:tPrec}), although due to computational limitations the simulation could only be run for $\simeq 25\%$ of a precession period \cite{Fragile+07}; analogous results have been seen in other GRMHD simulations \cite{Teixeira+14}.  In contrast, other GRMHD simulations have seen that MHD turbulence may disrupt bending waves and prevent rigid-body precession \cite{Sorathia+13}, although this result has been disputed \cite{Nealon+16}. Very recent GRMHD simulations have observed rigid-body precession that slows due to viscous spreading of the disk \cite{Liska+17}.

\begin{figure}
{ \begin{center}
\includegraphics[scale=1.30]{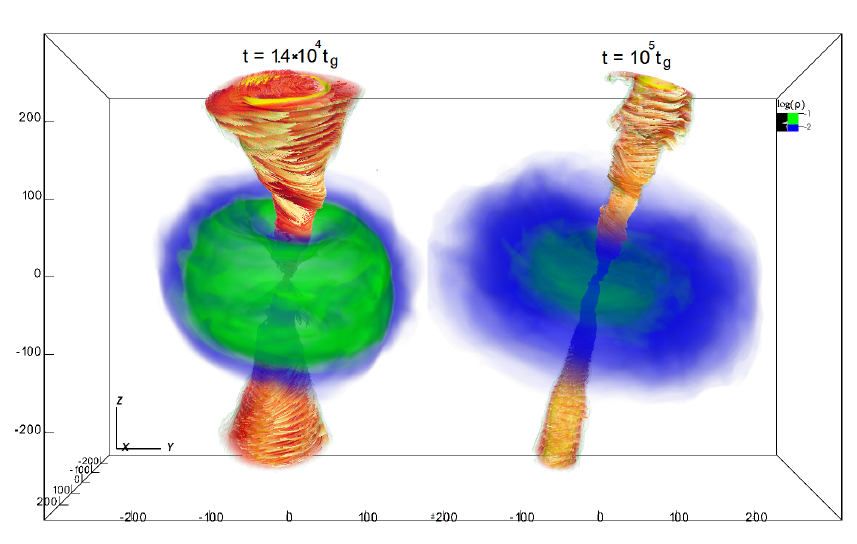}
\end{center}
\caption{\label{figure:JetPrecession}
Two snapshots taken at different times of GRMHD simulations of jet launching in a thick, compact accretion flow analogous to simple models of TDE disks.  The disk density is shown in green and blue, while the magnetic field lines threading the jet are shown in red and yellow.  In these simulations, both the disk and jet were seen to precess around the SMBH spin axis (the $z$-axis in this figure), as can be seen from the time-dependent jet orientation.  This figure was originally published as Fig.~2 of Liska {\it et al.} \cite{Liska+17}.
}  }
\end{figure}

Having summarized the theory of global precession in the tilted, thick accretion flows that are likely common in at least some TDEs (i.e. any efficiently circularizing subset of events), we can now ask whether global precession is likely to leave an observable imprint in TDE light curves.  As the inner disk precesses, an observer viewing the TDE from a constant line of sight will see a periodic or quasi-periodic modulation as he is exposed to time-varying projections of the inner disk annuli.  This is likely to manifest itself as a quasi-periodic oscillation in the thermal soft X-ray light curve \cite{Stone&Loeb12}.  If the disk precesses through an edge-on configuration, these oscillations could dramatically increase in amplitude.  As the observer loses visual contact with the innermost annuli, the observed emission will become much dimmer and redder \cite{Ulmer99}.  

If the precessing disk also launches a collimated, relativistic jet, even more dramatic signatures of disk precession may exist.  The jet (and associated hard X-ray emission) could sweep across the observer's line of sight in a lighthouse-like fashion, creating strong time variability in the hard X-ray signal \cite{Stone&Loeb12}.  Precession, and eventual alignment of the jet with the black hole spin axis, can also impart the relativistic jet with a complex and broad angular structure that could help explain the late-time radio re-brightening of Swift J1644+57 \cite{Tchekhovskoy+14}.  Of course, this hypothesis presupposes that jets launched from tilted accretion flows will follow the disk angular-momentum vector rather than the steady SMBH spin vector.  While the orientation of jets launched in tilted accretion flows was unclear for many years, recent GRMHD simulations have indicated that both jets and their parent disks precess about the SMBH spin axis, as is seen in Fig.~\ref{figure:JetPrecession} \cite{Liska+17}.

In summary, global precession seems likely to be present in the subset of TDEs that (i) rapidly circularize a large fraction of their debris and (ii) have an initial fallback rate $\dot{M}_{\rm peak} \gtrsim \dot{M}_{\rm Edd}$, although the fractional size of this subset is unclear.  Rigid-body precession will superimpose a quasi-periodic modulation on top of the secular trends observed in the thermal, soft X-ray light curves of TDE disks.  This modulation is likely of modest amplitude, but could become a dramatic, order-unity effect if the disk precesses through an edge-on configuration.  Even more dramatic precession effects could arise if the TDE launches a collimated jet which passes in and out of the observer's line of sight.  More theoretical work, and especially more GRMHD simulations, are needed to better understand the detailed evolution of a precessing thick disk, and how reliably any observable periodicity would encode the spin of the central SMBH.

\section{Observed TDE Flares}
\label{sec:obs}

Tidal disruption events have been the subject of theoretical predictions for almost half a century (\S \ref{sec:intro}, \S \ref{sec:basic}).  However, it was only more recently that TDEs passed from the realm of theoretical speculation into astronomical observation.  The first plausible TDE candidates were discovered in the 1990s via soft X-ray emission.  Since then, they have been seen across the electromagnetic spectrum, from radio to $\gamma$-ray wavelengths.  In this section, we will quickly review the current sample of TDE candidates, discuss our reasons for believing that the observed flares are, in fact, TDEs, and draw some preliminary connections to the relativistic effects discussed in this review.

\subsection{X-ray selected TDEs}

A common search strategy to find TDEs at X-ray wavelengths is to select new sources that present a large flux increase with respect to the upper limit from a previous observation of the same part of the sky. When the extragalactic nature of the new source can be confirmed (by identifying the host galaxy) and observations obtained over the following years show a roughly monotonic decline of flux, the event will be a strong TDE candidate. At this point, the only alternative scenario to explain this X-ray flare would be extreme variability from a pre-existing accretion disk in an AGN. Large X-ray flares from AGN are not uncommon, but fortunately most AGN can be identified by using observations at other wavelengths (e.g., via narrow or broad emission lines in the optical spectrum), or from archival X-ray observations. 

Since the detection of the first X-ray selected TDE candidates \citep{Bade96,KomossaBade99}, about a dozen similar flares from inactive (i.e., non-AGN) galaxies have been found (see e.g. the review of Ref. \cite{Komossa15})---although some of these X-ray sources have only very sparse follow-up observations, which limits our ability to assess the validity of the TDE identification \cite{Auchettl16}. A key feature of these X-ray selected TDEs is their generally thermal spectra, with very ``soft" blackbody temperatures ($T \sim 0.1$~keV). For most events, the typical luminosity is $\nu L_\nu \sim 10^{43}$~erg/s, implying blackbody radii consistent with emission from a few gravitational radii of the black hole.  Most of these X-ray selected flares also show light curves that decay as power-laws in time; some of the observed power-law indices are consistent with $-5/3$.  In short, the peak luminosities, thermal emission, and temporal evolution of the soft X-ray selected TDE sample are all in general qualitative agreement with the expected signatures of a compact, transient TDE accretion disk produced from efficient circularization of tidal debris (see \S \ref{sec:AccretionDiskEvolution}).

Besides the thermal X-ray TDEs, there is a class of TDE that show luminous ($\nu L_\nu \sim 10^{47}$\,erg\,s$^{-1}$) non-thermal emission  \citep{Bloom11,Levan11,Burrows11,Cenko11, Brown+17}. This emission is thought to originate from a collimated relativistic jet that was launched after the disruption, an interpretation supported by the detection of transient synchrotron radio emission \cite{Giannios&Metzger11} in follow-up observations \cite{Zauderer11}. When these sources were first discovered in 2011, the only plausible alternative explanation for their properties was an extreme type of stellar explosion \citep[i.e., the longest gamma-ray burst;][]{Quataert12}. But this alternate hypothesis is challenged by more recent follow-up observations \citep[e.g.,][]{Levan16,Kara16}.  

New jetted TDEs may be found by the {\it Swift} satellite and in radio-transient surveys \citep{vanVelzen12b,Donnarumma15a, Metzger+15}. New soft X-ray TDEs are currently found with a rate of about one per year by the XMM-slew survey \citep{Saxton17}. The detection rate of soft X-ray TDEs by the near-future eROSITA mission could be as high as $\sim 10^3$ per year \citep{Khabibullin14a}, and the proposed Einstein Probe would likely find $\sim100$ TDEs per year \cite{2015arXiv150607735Y}.

\subsection{Optical/UV-selected TDEs}

Three TDEs have been found using UV observations by the GALEX satellite \citep{Gezari06,Gezari09}. As in soft X-ray searches, the requirement for UV selection is a large flux increase from a galaxy that shows no signs of harboring an AGN. An advantage of UV selection is the lower amplitude of AGN variability compared to that at  X-ray wavelengths \citep{Gezari15}. In addition, most supernovae are UV faint, making UV observations a powerful tool for distinguishing TDEs from other transients.
The proposed UV satellite ULTRASAT could potentially find $\sim 100$ TDEs per year \cite{2014AJ....147...79S}, but for the near future we will have to rely on optical surveys to find new TDEs. 

The first TDEs found in an optical imaging survey were presented by Ref. \cite{vanVelzen10}.  After a systematic SDSS search of flares from galactic nuclei, this work identified two events that showed two distinct features separating them from nuclear supernovae: (i) a high blackbody temperature\footnote{Here we refer to ``high blackbody temperature'' only in comparison to supernova explosions; this blackbody temperature is of course quite low in comparison to that of the X-ray selected TDE sample.} ($T\approx 2\times 10^4$~K) that (ii) did not evolve with time. The previously known UV-selected TDEs, as well as new TDEs found in time-domain optical surveys such as Pan-STARRS \citep{Gezari12}, PTF \cite{Arcavi14}, ASAS-SN \cite{Holoien14}, and iPTF \cite{Blagorodnova+17}, also show these two features. Spectroscopic observations of optically selected TDEs reveal another common property \citep{Arcavi14}: helium and/or hydrogen emission lines that are broadened with a velocity of $\sim 10^4$ km/s. 

More than a dozen optical/UV selected TDE candidates have been found to date (see Refs. \cite{2018ApJ...852...72V, Hung+17} for recent compilations). The blackbody radii of these events are $\sim 10^3$ gravitational radii, much larger than the tidal radius, or the observed sizes of X-ray emitting regions. As such, the optical/UV properties of TDEs are different from what would be expected based on the basic physics of the disruption (cf. \S \ref{sec:NewtonianCircularization}). 

While several scenarios have been proposed to explain how TDEs could be responsible for the observed optical flares (see \S \ref{sec:AccretionDiskEvolution}), some authors have instead argued for different explanations. Accretion induced by collisions of stars on bound circular orbits (EMRIs) around the black hole \citep{Metzger17} or instabilities in a hitherto dormant accretion disk \citep{2018MNRAS.474.3307S} could potentially explain (some) of the optically selected TDEs. An important argument against these ``TDE-impostor" scenarios is the detection of a strong suppression in the optical/UV TDE detection rate at high SMBH masses \citep{2018ApJ...852...72V}; this lack of flares above the Hills mass predicted by Eq.~(\ref{eq:MHills}) is strong evidence that our current sample of optical/UV-selected tidal disruption flares are indeed due to genuine TDEs. 

The detection rate of TDE flares in optical surveys has been a few per year for most of the 2010s. This rate will soon increase by a factor $\sim 10$ with the start of the Zwicky Transient Facility (ZTF) \cite{Hung18}, while the Large Synoptic Survey Telescope (LSST) will detect thousands of TDEs per year \cite{vanVelzen10}. 

\subsection{Observed Indications of GR}
\label{sec:ObsGR}

The current sample of TDE candidates is modest in size, but two recent discoveries may have found indications of the general relativistic Hills mass.  The first of these was the discovery of the TDE candidate ASASSN-15lh.  Upon discovery, this flare was identified as the brightest supernova ever observed \cite{Dong+16}; with a peak optical/UV luminosity $L \approx 2\times 10^{45}~{\rm erg~s}^{-1}$, ASASSN-15lh outshone even other examples of ``superluminous supernovae'' \cite{Gal-Yam12}. This interpretation was challenged by Ref. \cite{Leloudas16}, which pointed out that the location of the transient in the nucleus of a massive host galaxy, and the host's lack of star formation, are unprecedented for superluminous supernovae, which typically occur in dwarf galaxies with high specific star formation rates.  These host properties are more naturally explained if ASASSN-15lh is due to a TDE, with one major caveat: the host galaxy's SMBH mass, if inferred from galaxy scaling relations, is uncomfortably close to the relativistic Hills limit.

\begin{figure}
{ \begin{center}
\includegraphics[scale=1.30]{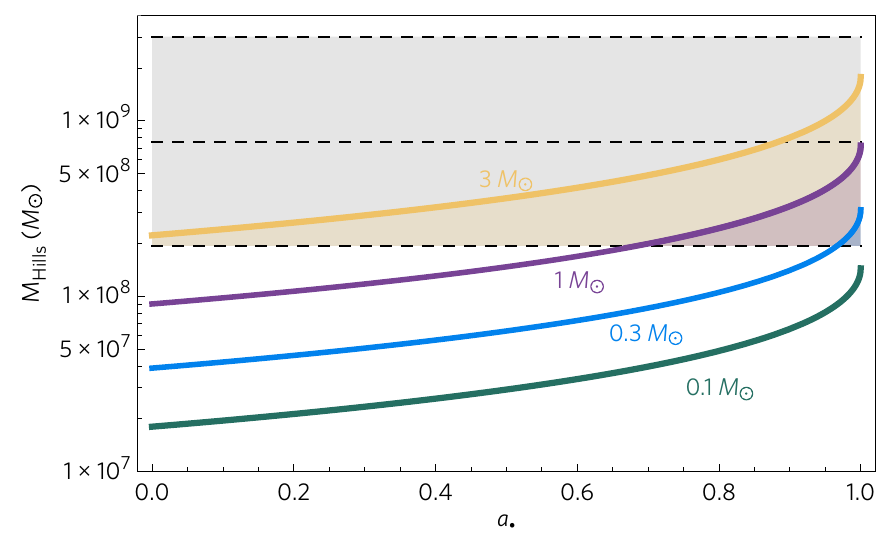}
\end{center}
\caption{\label{figure:MHills15lh}
The general relativistic Hills mass for prograde equatorial orbits as a function of dimensionless SMBH spin.  The green, blue, purple, and yellow curves show theoretical predictions for stars of mass $0.1M_\odot$, $0.3M_\odot$, $1M_\odot$, and $3M_\odot$, respectively.  The shaded gray region corresponds to one-sigma uncertainties on the mass of the SMBH in the host galaxy of ASASSN-15lh.  Colored shaded regions are combinations of SMBH spin and stellar mass that are compatible with the TDE interpretation of ASASSN-15lh.  This figure taken was originally published as Fig.~4 of Leloudas {\it et al.} \cite{Leloudas16}.
}  }
\end{figure}

The most recent analysis of the host galaxy of ASASSN-15lh \citep{Kruhler17} inferred a SMBH mass $M_\bullet \simeq 10^{8.3} M_\odot$. This exceeds the Newtonian Hills mass of Eq.~(\ref{eq:MHills}) for a TDE of a Solar-type star\footnote{Tidal disruption of much more massive stars, which have larger Hills masses, is disfavored both by their short lifetimes and the paucity of star formation in the ASASSN-15lh host.}, but not the relativistic Hills mass $M_H(a, \iota)$ for a star on a low-inclination orbit ($\iota \simeq 0$) tidally disrupted by a rapidly spinning Kerr black hole ($a \simeq M_\bullet$).  Fig.~\ref{figure:MHills15lh} illustrates combinations of SMBH spin and stellar mass that will lead to an electromagnetically luminous disruption.  This Kerr TDE interpretation may also explain the extreme optical luminosity of this flare, which is almost an order of magnitude higher at peak than that of other optically selected TDEs.  By requiring the disruption to occur on a prograde equatorial orbit around a SMBH with large spin, high radiative efficiency $\eta$ is favored, as is rapid debris circularization due to stream collisions promoted by apsidal precession (as described in \S \ref{sec:circularization}).  Furthermore, for this encounter, the debris returns to the black hole at a sub-Eddington rate, supporting the case for radiatively efficient accretion.  While the TDE interpretation is not undisputed \cite{Godoy-Rivera17}, and the unusual double-peaked light curve resembles neither known superluminous supernovae nor other TDE candidates\footnote{A further challenge is the short rise time of the flare, which appears much shorter than the predicted $t_{\rm min}$ for such a large SMBH.}, a detection of X-ray emission in follow-up observation \cite{Margutti17} presents a severe challenge to a stellar origin for this transient.

The high luminosity of ASASSN-15lh implies optical transients of this kind can be detected to great distance.  Consequently, the volumetric rate of these events must be orders of magnitude lower than the rate of other TDEs from lower-mass host galaxies. This is consistent with the interpretation that special conditions are required for SMBHs in such large galaxies to tidally disrupt stars (i.e., large spins and roughly prograde equatorial orbits).  Indeed, the very low volumetric rate of flares like ASASSN-15lh points to the second extant signature of the Hills mass: the statistical properties of our current sample of TDEs.  

\begin{figure}
{ \begin{center}
\includegraphics[scale=1.30]{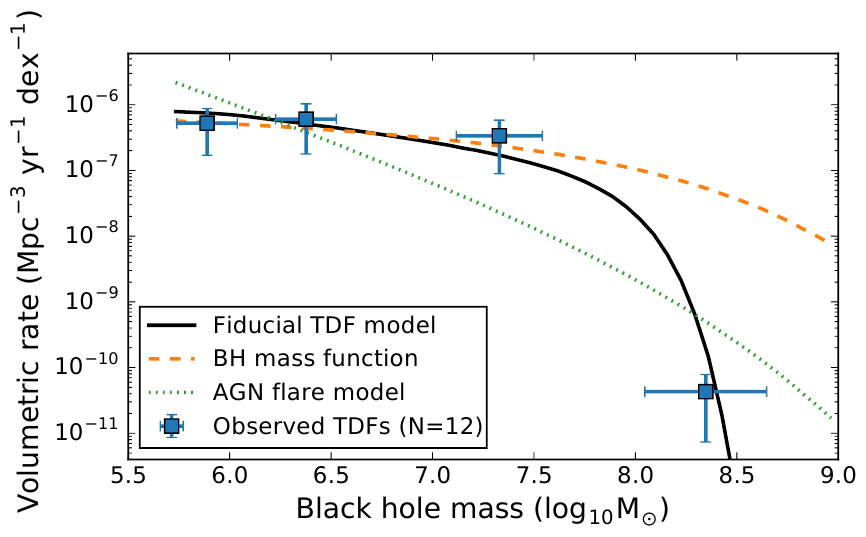}
\end{center}
\caption{\label{figure:SjoertHillsMass}
Volumetric TDE rates as functions of SMBH mass $M_\bullet$.  The blue data points represent a sample of 12 optically selected TDEs whose volumetric rates have been weighted by survey selection effects and flare luminosities.  The observed data clearly do not follow the volumetric SMBH mass distribution (dashed orange curve), nor do they follow a $1/M_\bullet$ scaling that is expected for AGN flares  (dotted green curve).  The super-exponential rate suppression for $M_\bullet \gtrsim 10^8 M_\odot$ appears to be a signature of the Hills mass seen in Fig.~\ref{F:supp} \cite{2012PhRvD..85b4037K}.  Note the caveat that the farthest right data point is the single, somewhat controversial TDE candidate ASASSN-15lh.  This figure was originally published as the upper panel of Fig.~3 of van Velzen \cite{2018ApJ...852...72V}.
}  }
\end{figure}

While the volumetric (or per-galaxy) rates of TDEs in the Universe are still debated \cite{Stone&Metzger16}, a recent effort to combine optically selected TDEs from a variety of time-domain surveys has estimated a volumetric mass function that shows a clear imprint of the Hills mass \cite{2018ApJ...852...72V}.  We illustrate this in Fig.~\ref{figure:SjoertHillsMass}, where the predicted super-exponential cutoff in the volumetric TDE rate for $M_\bullet \gtrsim M_H(a, \iota = \pi)$ shown in Fig.~\ref{F:supp} is indeed seen around $M_\bullet \simeq 10^8 M_\odot$.  Even if one remains agnostic about the TDE interpretation of ASASSN-15lh, the complete lack of flares from galaxies that host SMBHs more massive than $M_\bullet >10^{7.3} M_\odot$ is suggestive of rate suppression due to direct capture beyond the event horizons of the most massive black holes.

In summary, we have seen signs of the Hills mass - a fundamentally general-relativistic prediction - even in our limited, present-day TDE sample.  An exciting challenge for the coming decade will be to leverage the $\sim 10^3$ new TDEs likely to be seen by optical and X-ray time-domain surveys to better populate diagrams like Fig.~\ref{figure:SjoertHillsMass}.  By carefully accounting for the astrophysical and survey selection effects that go into this type of diagram, astronomers may obtain powerful constraints on the cosmic spin distribution of SMBHs with $10^{7.5} M_\odot \lesssim M_\bullet \lesssim 10^{8.5} M_\odot$.

\section{Discussion}
\label{sec:disc}

Tidal disruption events are a relatively novel class of astrophysical transients, and one with great potential for probing SMBH demography.  They also appear to represent a rare astrophysical example of fundamentally general-relativistic events, in that GR is not a first- or second-order correction to basically Newtonian dynamics, but rather the zeroth-order driver of qualitative observables (e.g. the simple question of whether electromagnetic emission can be produced at all, and more complex questions about the pace and efficiency of accretion-disk formation).  In this review, we have outlined the basic physical processes underlying tidal disruption, covering both Newtonian (\S \ref{sec:basic}) and general-relativistic (\S \ref{sec:GR1}) gravity.  The physics of disruption is reasonably well understood, so we have provided both simple analytic formulas and a survey of numerical hydrodynamic simulations of the disruption process.  The clearest imprint of GR (and black-hole spin) on the disruption process is the limiting Hills mass $M_H \sim 10^8 M_\odot$ above which electromagnetically luminous TDEs do not occur (i.e. the victim star is swallowed whole).  Other signatures of GR are also imprinted into the stellar debris during tidal disruption, but these are either more subtle or more speculative.

The aftermath of tidal disruption, however, is not so well understood.  Here there are many open theoretical questions, which are primarily but not exclusively of a hydrodynamical nature.  The process by which the dynamically cold debris streams from a TDE circularize into a luminous accretion flow (or fail to) remains quite uncertain, although it is increasingly clear that relativistic precession is critical for TDEs which do circularize efficiently.  We have summarized the hydrodynamical evolution of tidal-debris streams, as well as any accretion flows they may produce.  As with the dynamics of disruption, we have examined post-disruption hydrodynamics in both Newtonian (\S \ref{sec:NewtonianAccretion}) and relativistic (\S \ref{sec:GR2}) frameworks.  In our admittedly subjective judgment, the most important open questions in the theory of general-relativistic\footnote{There are other interesting questions in the field of TDEs as well, such as the apparent preference of TDEs for rare post-starburst host galaxies \cite{Arcavi14, French16, Stone+17}, but these have less bearing on the relativistic focus of this review.} tidal disruption seem to be:
\begin{enumerate}
    \item Under what conditions (if any) does the debris from a tidally shredded star settle into a quasi-circular accretion flow around the SMBH?
    \item How is the optical emission produced in optically luminous TDEs?  Is it powered primarily by circularization luminosity, or instead by reprocessing of X-ray/EUV photons from an inner accretion disk?  If the latter, what is the nature of the reprocessing material?
    \item What fraction of TDEs launch powerful relativistic jets, and what special conditions are required for this to occur?
\end{enumerate}
Answering these questions from first principles has been challenging due to the inherent dynamic range of the TDE problem (\S \ref{S:NumChal}), which has so far frustrated fully self-consistent hydrodynamical simulations.  Nonetheless, future progress - which will likely come from a mixture of theory and observation - is necessary if we hope to settle these questions and make predictive models for TDE emission, as is necessary if TDEs are to live up to their potential as probes of SMBH demographics.

At the moment, the most promising avenues for observing spin-related effects in TDE flares (or in statistical populations thereof) appear to be:
\begin{enumerate}
    \item The Hills mass.  The spin distribution of SMBHs in the mass range $10^{7.5} M_\odot \lesssim M_\bullet \lesssim 10^{8.5} M_\odot$ will determine the precise shape of the super-exponential cutoff in the {\it observed mass distribution} of TDE hosts.  Individual flares from SMBHs with $M_\bullet \gtrsim 10^8 M_\odot$ may require a minimum spin to be consistently interpreted as TDEs; ASASSN-15lh is the first (though still controversial) example of such a flare.
    \item Circularization delays from frame-dragging.  The early-time (rising) phase of a TDE light curve will be heavily affected by the efficiency of debris circularization, and much existing theoretical work (both analytic and numerical) indicates that strong spin-orbit misalignment in TDEs can lead to severe circularization delays.  This delay could be seen in the light curve of an individual TDE flare, or perhaps statistically, as the SMBH spin distribution will affect the TDE luminosity function.
    \item Global precession of efficiently circularized disks.  Periodic or quasi-periodic modulations of the thermal soft X-ray light curve may result from global precession of a quasi-circular accretion disk, in whatever subset of TDEs do circularize efficiently.  If the TDE launches a relativistic jet, a dramatic ``lighthouse-like'' effect could be seen in the hard X-ray light curve produced by the jet.
\end{enumerate}
Of course, the open theoretical questions concerning TDE hydrodynamics make it challenging to search for the second and third of these signatures. It is also possible that - even given perfectly predictive theoretical TDE models - systematic uncertainties or nuisance parameters may overwhelm some of these effects.  The only one which seems certain to manifest itself is the statistical, spin-dependent imprint of the Hills mass in populations of TDEs.  On the other hand, some of the more speculative applications of GR to TDE physics we have highlighted (e.g. thermonuclear fusion, shock breakout, or GWs from relativistic multiple compressions) may bear unexpected fruit.

While the primary focus of this review is theoretical, there is burgeoning observational interest in TDEs, as several dozen strong candidate flares have been discovered in the last two decades.  The advent of optical time-domain astronomy has greatly facilitated the search for TDE flares, and upcoming advances in this field (in particular, the ZTF and LSST surveys) are poised to expand our sample of optically bright events by orders of magnitude.  After its launch, the eROSITA instrument will likely play a similar role for X-ray bright TDEs.  We have provided a brief summary of the currently observed TDE sample (\S \ref{sec:obs}) and encourage interested readers to consult dedicated observational reviews and population analyses for more details \cite{Komossa15, Auchettl16, Hung17}.

It is clear that GR plays a leading role in multiple areas of tidal-disruption dynamics and hydrodynamics, but much more progress is needed to develop predictive theoretical models for TDE emission.  We have presented the open questions we view as most important in the hope that this review will serve as not just as a reference for known results, but also as a guide for researchers working on this promising frontier in relativistic astrophysics.

\section*{Acknowledgments}
Financial support was provided to NCS by NASA through Einstein Postdoctoral Fellowship Award Number PF5-160145.  MK is supported by the Alfred P. Sloan Foundation Grant No. FG-2015-65299 and NSF Grant No. PHY-1607031.  The work of RMC was funded by a Nicholas C. Metropolis Postdoctoral Fellowship under the auspices of the U.S. Dept. of Energy, and supported by its contract W-7405-ENG-36 to Los Alamos National Laboratory.

\bibliographystyle{spphys}
\bibliography{bib,general_desk}

\end{document}